\documentclass[aps,prb,twocolumn,superscriptaddress,noeprint,longbibliography]{revtex4-2}

\usepackage{amsmath,amssymb,bbold,mathtools}
\usepackage{graphicx,epstopdf}
\usepackage{dcolumn}
\usepackage{verbatim, float}
\usepackage{xcolor}
\usepackage[colorlinks=true]{hyperref}
\usepackage[normalem]{ulem}
\usepackage{enumitem}

\begin{document}
\title{Consistent Evaluation of Operators Involving the Position Operator in the Bloch Representation: Application to the Orbital Moment}
\author{Daehyeon An}
\email{daehyeon.an@kaist.ac.kr}
\affiliation{Physics Department, Korea Advanced Institute of Science and Technology}
\author{Junmo Jeon}
\email{junmojeon@sophia.ac.jp}
\affiliation{Physics Division, Sophia University}
\author{Se Kwon Kim}
\email{sekwonkim@kaist.ac.kr}
\affiliation{Physics Department, Korea Advanced Institute of Science and Technology}
\date{\today}

\begin{abstract}
The position operator plays a central role in condensed-matter observables such as velocity, orbital moment, and electric polarization.
In solid-state physics, the evaluation of operators incorporating the position operator has not reached a consensus, as observed in the operator-level discrepancy between the local circulation of Wannier functions and the self-rotation of wave packets.
Here, to achieve a consistent evaluation of such operators, we propose three rules for evaluating operators involving the position operator in the Bloch representation.
The rules are devised to satisfy physical conditions: independence from the choice of unit cell, preservation of Hermitian conjugacy for the product of operators, and recovery of the correct intraband velocity.
We further address the gauge dependence of the position operator and introduce a scheme termed gauge filtration, which systematically removes gauge-dependent contributions from the operators containing the position operator.
This methodology ensures that the quantities obtained from the operator evaluation correspond to observable physical phenomena.
By applying our framework, we reconcile the results concerning the self-rotation of the wave packet and the local circulation of the Wannier function.
We expect our proposal to establish a consistent framework for evaluating operators involving the position operator.
\end{abstract}
\maketitle

\section{Introduction}\label{ASection 1}
In solid-state physics, the position operator appears in the expressions of various physical quantities, including temperature-gradient potentials~\cite{Luttinger1964PhysRev.135.A1505}, electromagnetic perturbations~\cite{Adams1957PhysRev.107.698, Blount1962PhysRev.126.1636, Karplus1954PhysRev.95.1154, Zak1968PhysRev.168.686, Zak1969PhysRev.177.1151, Hasegawa1969PhysRev.177.1392, Kohn1959PhysRev.115.1460, Adams1959JPhysChemSolids.10.286, Roth1962JPhysChemSolids.23.433,Thouless1982PhysRevLett.49.405}, nonlinear optical processes involving electric polarization~\cite{Morimoto2021PhysRevB.104.075139, Boyd2020Book}, the quantum metric~\cite{Provost1980CommunMathPhys.76.289, Anandan1990PhysRevLett.65.1697, Mera2022PhysRevB.106.165133, Avdoshkin2023PhysRevB.107.245136, Piechon2016PhysRevB.94.134423, Ozawa2021PhysRevB.104.045103, Xu2025ArXiv.11425}, modern theories of polarization and orbital magnetization~\cite{Vanderbilt1993PhysRevB.48.4442, Thonhauser2011IntJModPhysB25.1429, Souza2004PhysRevB.69.085106, Resta2005ChemPhysChem.6.1815, Thonhauser2005PhysRevLett.95.137205, Ceresoli2006PhysRevB.74.024408, Ceresoli2010PhysRevB.81.060409, Aryasetiawan2019JPhysChemSolid.128.87}, and orbitronics~\cite{Bhowal2021PhysRevB.103.195309, Pezo2022PhysRevB.106.104414, Go2024NanoLett.24.5968, An2025PhysRevB.111.104436}.
However, evaluating the position operator within the Bloch representation is known to possess subtle issues.

The position operator exhibits a gauge dependence in infinite periodic systems~\cite{Vanderbilt1993PhysRevB.48.4442}.
This is associated with its dependence on the choice of origin; namely, the identical position can be relabeled from $\boldsymbol{r}$ to $\boldsymbol{r}+\boldsymbol{R}$ by shifting the origin from $\mathcal{O}$ to $\mathcal{O}'=\mathcal{O}-\boldsymbol{R}$.
Operators incorporating the position operator, such as the orbital moment operator $\hat{\boldsymbol{m}}=(\hat{\boldsymbol{r}}\times\hat{\boldsymbol{v}}-\hat{\boldsymbol{v}}\times\hat{\boldsymbol{r}})/4$~\cite{Sundaram1999PhysRevB.59.14915, Thonhauser2005PhysRevLett.95.137205, Pezo2022PhysRevB.106.104414, Go2024NanoLett.24.5968, Atencia2024AdvPhysX.9.2371972, An2025PhysRevB.111.104436}, where the velocity operator is given by $\hat{\boldsymbol{v}}=(i\hbar)^{-1}\{\hat{\boldsymbol{r}},\hat{H}\}_{-}$ and $\{\cdot,\cdot\}_{-}$ denotes the commutator, inherit this gauge dependence, as an origin shift affects the orbital moment ($\boldsymbol{m}\rightarrow\boldsymbol{m}+\boldsymbol{R}\times\boldsymbol{v}/2$).
Careful manipulation of the position operator is necessary to obtain a physical, gauge-independent quantity.

As a direct consequence of this nuisance in the position-operator evaluation, conventional evaluations of the position operator manifest three primary technical challenges.
First, conventional treatments of the position operator exhibit an unphysical dependence on the choice of unit cell, as noted in Refs.~\cite{Blount1962SolidStatePhys.13.305, Si2025EPL.149.26001, Vanderbilt2018Book}.
Second, evaluating the intraband group velocity using this operator fails to straightforwardly reproduce its standard expression $\boldsymbol{v}_n=\hbar^{-1}\boldsymbol{\partial_{k}}E_{n}(\boldsymbol{k})$, as noted in Refs.~\cite{Resta1998PhysRevLett.80.1800, Si2025EPL.149.26001}, where $E_{n}(\boldsymbol{k})$ is the $n$-th band energy, $n$ is the band index, $\hbar\boldsymbol{k}$ is the crystal momentum, $\hbar$ is the reduced Planck constant, and $\boldsymbol{\partial_{k}}=\boldsymbol{\partial}/\boldsymbol{\partial k}$.
Lastly, some previous formulations violate the Hermitian conjugate relation for the product of the position operator $\hat{\boldsymbol{r}}$ and a non-commuting Hermitian operator $\hat{O}$, i.e., $\langle \phi|\hat{\boldsymbol{r}}\hat{O}|\psi\rangle\neq \langle \psi|\hat{O}\hat{\boldsymbol{r}}|\phi\rangle^*$, where $^*$ denotes the complex conjugate, and $|\phi\rangle$ and $|\psi\rangle$ are Bloch states, as discussed in Ref.~\cite{Fuchs1940ProcRSocA.176.214, Bross1972ZPhysA.255.325} with the momentum operator, $\hat{O} =\hat{\boldsymbol{p}}$.
This violation results in an unphysical complex-valued expectation value for the orbital moment operator.

These underlying issues complicate the evaluation of the orbital moment operator, which is typically described via two distinct formulations: the local circulation of Wannier functions~\cite{Thonhauser2005PhysRevLett.95.137205, Ceresoli2006PhysRevB.74.024408} and the self-rotation of semiclassical wave packets~\cite{Xiao2005PhysRevLett.95.137204, Xiao2010RevModPhys.82.1959}.
Although conventional calculations often assert equivalence by yielding identical expressions for the magnetization~\cite{Thonhauser2011IntJModPhysB25.1429, Atencia2024AdvPhysX.9.2371972}, discrepancies arise that have rarely been analyzed explicitly when evaluating the orbital moment operator in the Wannier-function limit of the wave-packet distribution.
Establishing the complete equivalence between the two formalisms for the orbital magnetization at the operator level is crucial for understanding the intrinsic physical properties of the orbital moment operator and operators derived from it, such as the orbital moment current operator.

In this work, we propose three rules for evaluating operators involving the position operator that are contrived to satisfy the following physical conditions: independence from the choice of unit cell, preservation of Hermitian conjugacy for the product of operators, and recovery of the correct intraband velocity.
Furthermore, we demonstrate how to extract physically measurable quantities by systematically removing the gauge dependence introduced by the position operator.
We term this gauge-removal process \textit{gauge filtration}.
Applying these three rules, we establish the exact equivalence between the two formalisms for the orbital magnetization, the self-rotation of wave packets~\cite{Sundaram1999PhysRevB.59.14915, Xiao2005PhysRevLett.95.137204} and the local circulation of Wannier functions~\cite{Thonhauser2005PhysRevLett.95.137205, Ceresoli2006PhysRevB.74.024408} by explicitly deriving contributions previously omitted in intermediate steps.
We identify that gauge filtration is the appropriate gauge-removal process for the orbital moment operator by comparing theoretical routes that lead to gauge-independent orbital moment quantities.

The remainder of this paper is organized as follows.
In Sec.~\ref{ASection 2}, we establish the terminology and notation, and specify the physical constraints of the system under consideration.
Following a summary of our main results in Sec.~\ref{Sect: Summary}, we critically review previous representations of the position operator in the literature and identify their shortcomings in Sec.~\ref{ASection 3}.
In Sec.~\ref{ASection 4}, we propose three rules for evaluating operators to ensure calculational consistency of the position operator in the Bloch representation.
We then apply these rules to derive a general result for operators containing the position operator and verify that the outcome satisfies physical conditions in Sec.~\ref{ASection 5}.
In Sec.~\ref{BSection 2}, we discuss the origin of the gauge dependence of the position operator and its effects on operator evaluation.
To address this gauge dependence, Sec.~\ref{BSection 3} introduces the gauge filtration process, a systematic method for extracting physically measurable quantities.
In Sec.~\ref{BSection 4}, we apply gauge filtration to the orbital moment and clarify, within a unified operator framework, the exact equivalence between the local circulation~\cite{Thonhauser2005PhysRevLett.95.137205, Ceresoli2006PhysRevB.74.024408} and the self-rotation of wave packets~\cite{Sundaram1999PhysRevB.59.14915, Xiao2005PhysRevLett.95.137204}.
Finally, Sec.~\ref{Sect: Discussion} discusses the limitations of the proposed regularization framework, and Sec.~\ref{ASection 7} concludes the paper.

\section{Notation and terminology}\label{ASection 2}
Throughout this paper, operators are denoted by a hat (e.g., $\hat{O}$), boldface symbols represent vector quantities (e.g., $\boldsymbol{p}$), and plain fonts indicate scalars (e.g., $E_n$).
The dimensionality of the system is assumed to be arbitrary unless otherwise specified.
Accordingly, the dimensions of integrals and Dirac delta functions implicitly follow the dimensionality of their respective variables; for instance, $\int d\boldsymbol{k}$, $\int d\boldsymbol{r}$, and $\delta(\boldsymbol{k}'-\boldsymbol{k})$ correspond to $\int d^{d}k$, $\int d^{d}r$, and $\delta^{(d)}(\boldsymbol{k}'-\boldsymbol{k})$ respectively, in a general $d$-dimensional system.
For one-dimensional cases, they are denoted by $\int dk$, $\int dr$, and $\delta(k'-k)$.

In this paper, we consider Bloch states $\psi_{n\boldsymbol{k}}(\boldsymbol{r})=\langle \boldsymbol{r}|\psi_{n\boldsymbol{k}}\rangle=e^{i\boldsymbol{k}\cdot\boldsymbol{r}}u_{n\boldsymbol{k}}(\boldsymbol{r})=\langle \boldsymbol{r}|e^{i\boldsymbol{k}\cdot\hat{\boldsymbol{r}}}|u_{n\boldsymbol{k}}\rangle$ as the eigenstates of the Hamiltonian $\hat{H}$, where $|u_{n\boldsymbol{k}}\rangle$ denotes the cell-periodic part of the Bloch state~\cite{Vanderbilt2018Book}.
The inner product involving an operator implies integration over the position basis; for example,
\begin{equation}
\begin{split}
    \langle \psi_{n'\boldsymbol{k}'}| \hat{O}|\psi_{n\boldsymbol{k}}\rangle
    &=\int d\boldsymbol{r}\langle \psi_{n'\boldsymbol{k}'}|\boldsymbol{r} \rangle\langle \boldsymbol{r}|\hat{O}|\psi_{n\boldsymbol{k}}\rangle,\\
    \langle \psi_{n'\boldsymbol{k}'}| \hat{O}|\psi_{n\boldsymbol{k}}\rangle_\text{cell}
    &=\int_\text{cell} d\boldsymbol{r}\langle \psi_{n'\boldsymbol{k}'}|\boldsymbol{r} \rangle\langle \boldsymbol{r}|\hat{O}|\psi_{n\boldsymbol{k}}\rangle,\\
    \langle u_{n'\boldsymbol{k}'}| \hat{O}|u_{n\boldsymbol{k}}\rangle
    &=\int_\text{cell} d\boldsymbol{r}\langle u_{n'\boldsymbol{k}'}|\boldsymbol{r} \rangle\langle \boldsymbol{r}| \hat{O}|u_{n\boldsymbol{k}}\rangle.
\end{split}
\end{equation}

We adopt the following conventions for the integration range.
For $\langle \psi_{n'\boldsymbol{k}'}| \hat{O}|\psi_{n\boldsymbol{k}}\rangle$ and $\int d\boldsymbol{r}$, the integration domain is the entire continuous position space.
The integration domain restricts to a single unit cell when dealing with the cell-periodic part of the Bloch state, $\langle u_{n'\boldsymbol{k}'}| \hat{O}|u_{n\boldsymbol{k}}\rangle$, or when explicitly specified using a subscript, as in $\langle \psi_{n'\boldsymbol{k}'}| \hat{O}|\psi_{n\boldsymbol{k}}\rangle_\text{cell}$ and $\int_\text{cell} d\boldsymbol{r}$.
All integrations in the reciprocal space ($\int d\boldsymbol{k}$) are performed over the first Brillouin zone, even when the subscript $\text{B.Z.}$ is omitted.

Square brackets are employed to denote the matrix element of an operator in the Bloch representation, indexed by subscripts: 
$[O(\boldsymbol{k})]_{n'n}=\langle u_{n'\boldsymbol{k}}|\hat{O}(\boldsymbol{k})|u_{n\boldsymbol{k}}\rangle=\langle\psi_{n'\boldsymbol{k}}|\hat{O}|\psi_{n\boldsymbol{k}}\rangle_{\text{cell}}$.
Similarly, the Berry connection matrix is defined as $[A(\boldsymbol{k})]_{n'n}=\langle u_{n'\boldsymbol{k}}|i\boldsymbol{\partial_{k}} u_{n\boldsymbol{k}}\rangle=\int_\text{cell} d\boldsymbol{r} u_{n'\boldsymbol{k}}^*(\boldsymbol{r})i\boldsymbol{\partial_{k}}u_{n\boldsymbol{k}}(\boldsymbol{r})$.
A product chain of square brackets represents matrix multiplication, where the matrix indices are denoted by subscripts on the outermost square brackets; that is, $\big[[\cdot][\cdot]\big]_{n'n}=\sum_{m}[\cdot]_{n'm}[\cdot]_{mn}$. 

This work focuses on spinless, non-degenerate, and infinitely periodic systems with mutually orthogonal primitive vectors, such as rectangular or orthorhombic lattices.
The Hermitian operators under consideration include the position operator and any operator $\hat{O}$ possessing the following properties:
\begin{enumerate}
    \item First, an operator $\hat{O}$ is local if its matrix elements in the position basis are supported only at coincident points.
    More generally, in $d$ dimensions, it can be represented as $\langle \boldsymbol{r}'|\hat{O}| \boldsymbol{r}\rangle = \sum_{n}[\prod_{i=1}^{d}O_{n;n_i}(\boldsymbol{r}) ( \partial^{i}_{r'})^{n_i}] \delta^{(d)}( \boldsymbol{r}' - \boldsymbol{r})$, where $n=\sum_{i=1}^{d} n_i$; i.e., as a linear combination of derivatives of the Dirac delta function with coefficient functions $O_{n;n_i}(\boldsymbol{r})$.
     This form encompasses multiplicative operators containing only $n=0$ terms as well as differential operators, such as the momentum operator.
    \item Second, an operator $\hat{O}$ is cell-periodic if it respects the discrete translational symmetry of the crystal lattice, $O(\boldsymbol{r})=O(\boldsymbol{r}+\boldsymbol{R})$ for any Bravais lattice vector $\boldsymbol{R}$.
\end{enumerate}
All operators denoted by $\hat{O}$ and $\hat{Q}$ treated in this study, with the exception of the position operator $\hat{\boldsymbol{r}}$, are cell-periodic local operators.
Typical examples, when represented in the cell-periodic Bloch basis, include the momentum operator $\hat{\boldsymbol{p}}$ and the Hamiltonian operator $\hat{H}=\hat{\boldsymbol{p}}^2/2m+V(\hat{\boldsymbol{r}})$, where $m$ is the mass of the particle and $V(\hat{\boldsymbol{r}})=V(\hat{\boldsymbol{r}}+\boldsymbol{R})$ is the cell-periodic potential.
We define a product chain of operators as a sequence in the form $\hat{Q}_1\hat{Q}_2\cdots \hat{Q}_n$, where each $\hat{Q}_i$ is an operator; furthermore, a composite operator is defined as the sum of such chains acting as a single operator.

\section{Summary of main results}\label{Sect: Summary}
We propose a regularization scheme to obtain consistent evaluations of operators involving the position operator, alongside a procedure to remove the ensuing gauge dependence.
Under this regularization scheme, the evaluation of the band-diagonal matrix elements for a product chain of operators $\hat{O}\hat{\boldsymbol{r}}\hat{Q}$ yields the following:
\begin{widetext}
\begin{equation}
\begin{split}
    \lim_{N\rightarrow\infty}\frac{1}{N}\langle \Psi^{N}_{n'\boldsymbol{k}}|\hat{O}\hat{\boldsymbol{r}}\hat{Q}|\Psi^{N}_{n\boldsymbol{k}}\rangle
    =\big[[O(\boldsymbol{k})][\boldsymbol{A}(\boldsymbol{k})][Q(\boldsymbol{k})]\big]_{n'n}
    +\sum_{m}\frac{1}{2}\Big([O(\boldsymbol{k})]_{n'm}i\boldsymbol{\partial_{k}}[Q(\boldsymbol{k})]_{mn}
    -i\boldsymbol{\partial_{k}}[O(\boldsymbol{k})]_{n'm}[Q(\boldsymbol{k})]_{mn}\Big).
\end{split}
\end{equation}
Here, $N$ denotes the number of unit cells of the system, and $|\Psi^{N}_{n\boldsymbol{k}}\rangle$ is a $\Delta_N$-regularized Bloch state defined by $|\Psi^{N}_{n\boldsymbol{k}}\rangle=\int d\boldsymbol{k}''\Delta_{N}(\boldsymbol{k}-\boldsymbol{k}'')|\psi_{n\boldsymbol{k}''}\rangle$ with distribution kernel $\Delta_{N}(\boldsymbol{k})$.
Applying this framework, the orbital moment operator $\hat{\boldsymbol{m}}=(\hat{\boldsymbol{r}}\times\hat{\boldsymbol{v}}-\hat{\boldsymbol{v}}\times\hat{\boldsymbol{r}})/4$ is evaluated as:
\begin{equation}
\begin{split}
    \lim_{N\rightarrow\infty}\frac{1}{N}\langle \Psi^{N}_{n\boldsymbol{k}}|\hat{\boldsymbol{m}}|\Psi^{N}_{n\boldsymbol{k}}\rangle
    =\frac{1}{2i\hbar}\Big(
    [\boldsymbol{A}]_{nn}\times i\boldsymbol{\partial_{k}}E_{n}
    +\sum_{m}\big(E_{n}-E_{m}\big)
    [\boldsymbol{A}]_{nm}\times[\boldsymbol{A}]_{mn}\Big).
\end{split}
\end{equation}
\end{widetext}
The integral of this result over the Brillouin zone---multiplied by a factor of two to ensure exact correspondence with the orbital moment convention---is identical at the operator level to the results obtained within both the self-rotation of wave packets in the Wannier-function limit and the local circulation in the Wannier representation, provided that previously omitted implicit contributions that are discussed below in Sec.~\ref{BSection 4} are explicitly included in the integration.
Due to the inherent gauge dependence of the position operator, a gauge-removal process, termed gauge filtration, is necessary to obtain physically meaningful results.
Applying gauge filtration to the evaluated orbital moment operator extracts the following gauge-invariant result:
\begin{equation}
\begin{split}
    &\big([\boldsymbol{m}(\boldsymbol{k})]_{nn}\big)^{\text{G.F.}}\\
    &\quad=\frac{1}{2i\hbar}\langle\boldsymbol{\partial_{k}} u_{n\boldsymbol{k}}|\times\big(E_{n}(\boldsymbol{k})-\hat{H}(\boldsymbol{k})\big)|\boldsymbol{\partial_{k}}u_{n\boldsymbol{k}}\rangle,
\end{split}
\end{equation}
where $\text{G.F.}$ stands for the gauge filtration.

\section{Issues regarding the position operator in the literature}\label{ASection 3}
In this section, we examine several issues associated with the position operator in various Bloch representations discussed in the existing literature~\cite{Karplus1954PhysRev.95.1154, Sundaram1999PhysRevB.59.14915, Souza2004PhysRevB.69.085106, Bhowal2021PhysRevB.103.195309, Moessner2021Book, Cohen2016Book, Shankar2017Book, Vanderbilt2018Book, Souza2004PhysRevB.69.085106, Pezo2022PhysRevB.106.104414, Cayssol2021JPhysMater.4.034007, Xu2025ArXiv.11425, Si2025EPL.149.26001}.

\subsection{Problems on the Hermitian conjugate relation of the operator}\label{Problematic position operator Case 1}
To begin our analysis, for a state $|g\rangle$ in the Bloch representation, $|g\rangle=\sum_{n}\int \frac{d\boldsymbol{k}}{V_{\text{B.Z.}}}|\psi_{n\boldsymbol{k}}\rangle g_{n}(\boldsymbol{k})$, where $V_{\text{B.Z.}}$ denotes the volume of the first Brillouin zone, the action of the position operator in position space is formally expressed as
\begin{equation}
\begin{split}
    \langle \boldsymbol{r}| \hat{\boldsymbol{r}}|g\rangle
    =&\sum_{n',n}\int \frac{d\boldsymbol{k}}{V_{\text{B.Z.}}}\psi_{n'\boldsymbol{k}}(\boldsymbol{r})\\
    &\times\Big(
    \delta_{n'n} i\boldsymbol{\partial_{k}}g_{n}(\boldsymbol{k})+[\boldsymbol{A}(\boldsymbol{k})]_{n'n}g_{n}(\boldsymbol{k})
    \Big).
\end{split}
\end{equation}
This formulation assumes that both $\psi_{n'\boldsymbol{k}}(\boldsymbol{r})$ and $g_{n}(\boldsymbol{k})$ are well-defined within the first Brillouin zone and remain free of any singularities~\footnote{
Ref.~\cite{Blount1962SolidStatePhys.13.305} effectively neglects surface terms by adopting a gauge in which the Bloch states are assumed to be smooth and single-valued across the Brillouin zone (i.e., avoiding functional discontinuities).
However, as discussed in Ref.~\cite{Vanderbilt2018Book}, it is generally impossible to define a single globally smooth and periodic Bloch gauge over the entire Brillouin zone for bands with nontrivial topology.
Consequently, this necessitates the introduction of multiple Berry-connection patches and resultant phase discontinuities at the boundaries between them.
Such discontinuities in the Berry connection may manifest as effective surface terms in reciprocal space associated with each patch boundary.
}.
In the cell-periodic Bloch representation, it is customary to effectively represent matrix elements of the position operator as $\hat{\boldsymbol{r}}\sim i\boldsymbol{\partial_{k}}$, e.g., $\hat{\boldsymbol{r}}|u_{n\boldsymbol{k}}\rangle g_{n}(\boldsymbol{k})=i\boldsymbol{\partial_{k}}\big(|u_{n\boldsymbol{k}}\rangle g_{n}(\boldsymbol{k})\big)$ or equivalently as $\hat{\boldsymbol{r}}=i\boldsymbol{\partial_{k}}\delta_{n'n}+[\boldsymbol{A}(\boldsymbol{k})]_{n'n}$, where $i\boldsymbol{\partial_{k}}$ acts on the factor in the cell-periodic Bloch basis $|u_{n\boldsymbol{k}}\rangle$~\cite{Cohen2016Book, Shankar2017Book, Moessner2021Book, Cayssol2021JPhysMater.4.034007, Pezo2022PhysRevB.106.104414, Souza2004PhysRevB.69.085106, Xu2025ArXiv.11425}.
However, a theoretical issue emerges regarding the term $i\boldsymbol{\partial_{k}}$: the intraband value becomes complex for anticommutator-based operators $\hat{\boldsymbol{Q}}^{\text{AC}}=\{\hat{O},\hat{\boldsymbol{r}}\}_{+}/2$, where $\{\cdot,\cdot\}_{+}$ is anticommutator, such as the $z$ component orbital moment operator, $\hat{m}^{z}=(\{\hat{v}^y, \hat{r}^x \}_{+}-\{\hat{v}^x,\hat{r}^y\}_{+})/4$.

From a strictly mathematical perspective~\cite{Juric2022Universe.8.129, Song2024ArXiv.02519}, it may be argued that the Hermiticity of a composite operator $\hat{\boldsymbol{Q}}^{\text{C}}=\{\hat{O},\hat{\boldsymbol{r}}\}_{-}/(2i)$---such as the velocity operator $\hat{\boldsymbol{v}} = i \{\hat{H}, \hat{\boldsymbol{r}}\}_{-} /\hbar$---is defined strictly for the expression as a whole.
Within this framework, individual terms such as $\hat{O}\hat{\boldsymbol{r}}$ do not necessarily need to satisfy the Hermitian conjugate relation $\langle\psi_{n'\boldsymbol{k}}|\hat{\boldsymbol{r}}\hat{O}|\psi_{n\boldsymbol{k}}\rangle=\langle\psi_{n\boldsymbol{k}}|\hat{O}\hat{\boldsymbol{r}}|\psi_{n'\boldsymbol{k}}\rangle^{*}$, primarily because the codomain of the position operator may no longer coincide with the domain of the operator $\hat{O}$.

Physically, we impose three expectations to ensure theoretical consistency.
These include the Hermiticity of the position operator itself, $\langle\psi_{n'\boldsymbol{k}}|\hat{\boldsymbol{r}}|\psi_{n\boldsymbol{k}}\rangle=\langle\psi_{n\boldsymbol{k}}|\hat{\boldsymbol{r}}|\psi_{n'\boldsymbol{k}}\rangle^{*}$, as well as the Hermiticity of both commutator-based operators $\hat{\boldsymbol{Q}}^{\text{C}}$ and anticommutator-based operators $\hat{\boldsymbol{Q}}^{\text{AC}}$, such as the velocity operator and orbital moment operator, respectively~\cite{Pezo2022PhysRevB.106.104414, Bhowal2021PhysRevB.103.195309, Kontani2009PhysRevLett.102.016601, Mund2011JApplPhys.110.073914, Ding2020PhysRevLett.125.177201, He2020NatCommun.11.1650, Choi2023Nature.619.7968}.
To satisfy these requirements, it is physically intuitive and methodologically advantageous to impose the Hermitian conjugate relation, even though this constraint is more stringent than the minimal mathematical requirement for the Hermiticity of the commutator alone.

Based on these expectations, for instance, the intraband velocity is expressed as $\langle u_{n\boldsymbol{k}}|\hat{\boldsymbol{v}}|u_{n\boldsymbol{k}}\rangle=(i\hbar)^{-1}\langle u_{n\boldsymbol{k}}|\hat{\boldsymbol{r}}\hat{H}(\boldsymbol{k})-\hat{H}(\boldsymbol{k})\hat{\boldsymbol{r}}|u_{n\boldsymbol{k}}\rangle=(i\hbar)^{-1}(C-C^{*})$, where $C=\langle u_{n\boldsymbol{k}}|\hat{\boldsymbol{r}}\hat{H}(\boldsymbol{k})|u_{n\boldsymbol{k}}\rangle$, and $\hat{H}(\boldsymbol{k})$ is the Hamiltonian operator in the $|u_{n\boldsymbol{k}}\rangle$ basis.
To ensure that the evaluation of the complex number $C$ preserves the Hermitian conjugate relation for the operator product involving the position operator, while also correctly reproducing the intraband velocity $\hbar^{-1}\boldsymbol{\partial_{k}}E_{n}(\boldsymbol{k})$, it becomes necessary to adopt the following expressions:
\begin{equation}\label{Expected position operator}
\begin{split}
    \langle u_{n\boldsymbol{k}}|\hat{\boldsymbol{r}}\hat{H}(\boldsymbol{k})|u_{n\boldsymbol{k}}\rangle
    &=[\boldsymbol{A}(\boldsymbol{k})]_{nn}E_{n}(\boldsymbol{k})+\frac{1}{2}i\boldsymbol{\partial_{k}}E_{n}(\boldsymbol{k}),\\
    \langle u_{n\boldsymbol{k}}|\hat{H}(\boldsymbol{k})\hat{\boldsymbol{r}}|u_{n\boldsymbol{k}}\rangle
    &=[\boldsymbol{A}(\boldsymbol{k})]_{nn}E_{n}(\boldsymbol{k})-\frac{1}{2}i\boldsymbol{\partial_{k}}E_{n}(\boldsymbol{k}).
\end{split}
\end{equation}
The evaluation result for the position operator in Eq.~\eqref{Expected position operator} is nontrivial; specifically, it cannot be obtained if the differential operator $i\boldsymbol{\partial_{k}}$ acts only in a fixed direction~\cite{Xu2025ArXiv.11425}.
It appears that such unidirectional action---either to the right or to the left---may unintentionally break the Hermitian conjugate relation, leading to a situation where the anticommutator-based operator $\hat{\boldsymbol{Q}}^{\text{AC}}$ acquires a complex-valued band-diagonal component.

\subsection{Breakdown of the Dirac delta function identity}\label{Problematic position operator Case 2}
When the function $f(x)$ is multiplied by a Dirac delta function $\delta(x-x')$, we can replace the variable of the function with the other variable in the Dirac delta function, by applying a naive identity: $f(x)\delta(x-x')=f(x')\delta(x-x')$.
This identity results in different gauge dependence when it is used to evaluate the position operator in the Bloch representation.

Without exploiting the identity, one well-recognized approach for evaluating the position operator~\cite{Blount1962SolidStatePhys.13.305} is given by
\begin{equation}\label{r in Bloch}
\begin{split}
    &\langle \psi_{n'\boldsymbol{k}'}|\hat{\boldsymbol{r}}|\psi_{n\boldsymbol{k}''}\rangle
    =V_\text{B.Z.}\Bigg(
    \int_\text{cell} d\boldsymbol{r} \delta(\boldsymbol{k}'-\boldsymbol{k}'')\\
    &\times e^{-i(\boldsymbol{k}'-\boldsymbol{k}'')\cdot \boldsymbol{r}}u_{n'\boldsymbol{k}'}^*(\boldsymbol{r})i\boldsymbol{\partial_{k''}}u_{n\boldsymbol{k}''}(\boldsymbol{r})\\
    &-i\boldsymbol{\partial_{k''}}\int_\text{cell} d\boldsymbol{r}
    \delta(\boldsymbol{k}'-\boldsymbol{k}'')\psi_{n'\boldsymbol{k}'}^*(\boldsymbol{r})\psi_{n\boldsymbol{k}''}(\boldsymbol{r})
    \Bigg),
\end{split}
\end{equation}
where $\psi_{n'\boldsymbol{k}'}^*(\boldsymbol{r})\psi_{n\boldsymbol{k}''}(\boldsymbol{r})=e^{-i(\boldsymbol{k}'-\boldsymbol{k}'')\cdot \boldsymbol{r}}u_{n'\boldsymbol{k}'}^*(\boldsymbol{r})u_{n\boldsymbol{k}''}(\boldsymbol{r})$ (see Appendix~\ref{Appendix A}).

In several studies~\cite{Karplus1954PhysRev.95.1154, Sundaram1999PhysRevB.59.14915, Souza2004PhysRevB.69.085106, Bhowal2021PhysRevB.103.195309, Esteve-Paredes2023SciPostPhysCore.6.1.002}, exploiting the identity has been utilized as an alternative evaluation of the position operator:
\begin{equation}\label{many researchers}
\begin{split}
    \langle \psi_{n'\boldsymbol{k}'}|\hat{\boldsymbol{r}}|\psi_{n\boldsymbol{k}''}\rangle
    &=V_\text{B.Z.}\Big(\delta(\boldsymbol{k}'-\boldsymbol{k}'')[\boldsymbol{A}(\boldsymbol{k}'')]_{n'n} \\
    &-i\boldsymbol{\partial_{k''}}\delta(\boldsymbol{k}'-\boldsymbol{k}'')\delta_{n'n}
    \Big).
\end{split}
\end{equation}
Remarkably, Eq.~\eqref{many researchers} exhibits distinct behavior compared to Eq.~\eqref{r in Bloch} under a gauge transformation $|\tilde{\psi}_{n\boldsymbol{k}}\rangle=e^{i\chi_{n}(\boldsymbol{k})}|\psi_{n\boldsymbol{k}}\rangle$.
Under this gauge transformation, the expression in Eq.~\eqref{many researchers} acquires an additional diagonal contribution, $-V_\text{B.Z.}\delta(\boldsymbol{k}'-\boldsymbol{k}'')\boldsymbol{\partial_{k''}}\chi_{n}(\boldsymbol{k}'')\delta_{n'n}$, from the Berry connection.
In contrast, the expression in Eq.~\eqref{r in Bloch} transforms as $\langle\tilde{\psi}_{n'\boldsymbol{k}'}|\hat{\boldsymbol{r}}|\tilde{\psi}_{n\boldsymbol{k}''}\rangle=e^{i(\chi_{n}(\boldsymbol{k}'')-\chi_{n'}(\boldsymbol{k}'))}\langle \psi_{n'\boldsymbol{k}'}|\hat{\boldsymbol{r}}|\psi_{n\boldsymbol{k}''}\rangle$.
This discrepancy demonstrates that the naive use of the identity---$\delta(\boldsymbol{k}'-\boldsymbol{k}'')\psi_{n'\boldsymbol{k}'}^*(\boldsymbol{r})=\delta(\boldsymbol{k}'-\boldsymbol{k}'')\psi_{n'\boldsymbol{k}''}^*(\boldsymbol{r})$~\cite{Song2024ArXiv.02519}---is no longer valid in the presence of derivatives acting on the Dirac delta function and on $\boldsymbol{k}$-dependent Bloch states.

\subsection{Dependence on the choice of unit cell}\label{Problematic position operator Case 3}
In principle, the choice of the unit cell should not alter the physical results when evaluating operators under periodic boundary conditions.
However, it has been observed that evaluating the position operator via a unit-cell-restricted integration, $\int_\text{cell} d\boldsymbol{r} u_{n'\boldsymbol{k}}^*(\boldsymbol{r})\boldsymbol{r}u_{n\boldsymbol{k}}(\boldsymbol{r})$, inherently introduces an explicit dependence on the chosen unit cell~\cite{Blount1962SolidStatePhys.13.305, Si2025EPL.149.26001, Vanderbilt2018Book}.
This leads to an unphysical situation in which distinct but equivalent choices of the unit cell yield inconsistent numerical values for the same system, indicating that such cell-dependent position matrix elements are physically ill-defined.

\section{Three rules for evaluating operators}\label{ASection 4}
To address the challenges identified in Sec.~\ref{ASection 3} regarding previous evaluations of the position operator, we propose three rules for evaluating an operator $\hat{O}$ in the Bloch representation, designed to satisfy the requisite physical conditions.
In this section, we explain the proposed rules summarized in Table~\ref{Tab: Rules}. 
\begin{table*}[htb!]
    \centering
    \begin{tabular}{||l|c||}
    \hline
         \multicolumn{2}{||c||}{Objective: Evaluate $\hat{O}$ between $\langle \psi_{n'\boldsymbol{k}'}|$ and $|\psi_{n\boldsymbol{k}}\rangle$}
         \\ \hline
         \quad Rule 1: Take the lattice average \quad & \quad $\lim_{N\rightarrow\infty}N^{-1}\langle \psi_{n'\boldsymbol{k}'}|\hat{O}|\psi_{n\boldsymbol{k}}\rangle$ \quad \\ \hline 
         \quad Rule 2: Use the $\Delta_N$-regularized state $|\Psi^{N}_{n\boldsymbol{k}}\rangle$ \quad & \quad $\lim_{N\rightarrow\infty}N^{-1}\langle \psi_{n'\boldsymbol{k}'}|\hat{O}|\psi_{n\boldsymbol{k}}\rangle\rightarrow\lim_{N\rightarrow\infty}N^{-1}\langle \Psi^{N}_{n'\boldsymbol{k}'}|\hat{O}|\Psi^{N}_{n\boldsymbol{k}}\rangle \quad $ \quad 
         \\ \hline 
         \quad Rule 3: Discretize $\boldsymbol{k}$ and take the thermodynamic limit \quad & \quad $\lim_{N\rightarrow\infty}N^{-1}\langle \Psi^{N}_{n'\boldsymbol{k}'}|\hat{O}|\Psi^{N}_{n\boldsymbol{k}}\rangle 
    \rightarrow [O(\boldsymbol{k})]_{n'n}\delta_{\boldsymbol{k}'\boldsymbol{k}}$ \quad 
    \\
    \hline
    \end{tabular}
    \caption{Summarized flow of applying rules for evaluating the operator $\hat{O}$ in the Bloch representation.}
    \label{Tab: Rules}
\end{table*}

The first rule prescribes the use of the lattice average to evaluate the matrix element of the operator $\hat{O}$, expressed as $\lim_{N\rightarrow\infty}N^{-1}\langle \psi_{n'\boldsymbol{k}'}|\hat{O}|\psi_{n\boldsymbol{k}}\rangle$.
This represents the matrix element averaged over the entire system of $N$ unit cells subject to the Born--von Karman boundary conditions.
This approach evaluates the operator across the entire system rather than restricting it to a specific unit cell; the inherent divergence of the evaluation is regularized by the factor $N$, which corresponds to the order of the divergence arising from the infinite number of unit cells in the thermodynamic limit.

The second rule suggests substituting the Bloch state $|\psi_{n\boldsymbol{k}}\rangle$ with the $\Delta_N$-regularized Bloch state $|\Psi^{N}_{n\boldsymbol{k}}\rangle$,
\begin{equation}\label{Reg Bloch function}
    |\Psi^{N}_{n\boldsymbol{k}}\rangle=\int d\boldsymbol{k}''\Delta_{N}(\boldsymbol{k}-\boldsymbol{k}'')|\psi_{n\boldsymbol{k}''}\rangle,
\end{equation}
which is considered as a superposition of Bloch states $|\psi_{n\boldsymbol{k}''}\rangle$ that converges to the exact Bloch state $|\psi_{n\boldsymbol{k}}\rangle$ in the thermodynamic limit.
To fulfill this role, the regularizing kernel $\Delta_{N}(\boldsymbol{k}-\boldsymbol{k}')$ must satisfy the following conditions: it must be physically well-defined on the Brillouin zone, reduce to $\frac{N}{V_{\text{B.Z.}}}\delta_{\boldsymbol{k}\boldsymbol{k}'}$ for a finite $N$, and approach the Dirac delta function $\delta(\boldsymbol{k}-\boldsymbol{k}')$ in the thermodynamic limit.
The provided kernel is sufficient to apply the proposed rules.

The third rule is introduced to regularize expressions that would otherwise involve the mathematically ill-defined naive product of two Dirac delta-like distributions originating from the $\Delta_{N}$ functions in the thermodynamic limit.
To resolve this, we prescribe a strict two-step treatment for the system size $N$: first, discretizing $\boldsymbol{k}$ for a finite system, and subsequently taking the thermodynamic limit.
Here, the thermodynamic limit is taken only after addressing the $N$ factors within the finite $N$ regime, i.e., the discrete $\boldsymbol{k}$ regime.

The first step discretizes the continuous variable $\boldsymbol{k}''$ in Eq.~\eqref{Reg Bloch function} to align it with the discrete $\boldsymbol{k}$, which characterizes the state $|\Psi^{N}_{n\boldsymbol{k}}\rangle$ for a Born--von Karman crystal comprising $N$ unit cells.
In the calculation of the operator $\hat{O}$,
\begin{equation}\label{Eval oper using third rule}
\begin{split}
    &\lim_{N\rightarrow\infty}\frac{1}{N}\langle \Psi^{N}_{n'\boldsymbol{k}'}|\hat{O}|\Psi^{N}_{n\boldsymbol{k}}\rangle 
    = \lim_{N\rightarrow\infty}\frac{V_{\text{B.Z.}}}{N}\int d\boldsymbol{k}''\\
    &\times\Delta_{N}(\boldsymbol{k}'-\boldsymbol{k}'')[O(\boldsymbol{k}'')]_{n'n}\Delta_{N}(\boldsymbol{k}-\boldsymbol{k}''),
\end{split}
\end{equation}
this discretization replaces $\Delta_{N}(\boldsymbol{k}-\boldsymbol{k}')$ and $\int \frac{d\boldsymbol{k}}{V_{\text{B.Z.}}}$ with $\frac{N}{V_{\text{B.Z.}}}\delta_{\boldsymbol{k}\boldsymbol{k}'}$ and $N^{-1}\sum_{\boldsymbol{k}}$, respectively, yielding $\lim_{N\rightarrow\infty}\sum_{\boldsymbol{k}''}[O(\boldsymbol{k}'')]_{n'n}\delta_{\boldsymbol{k}'\boldsymbol{k}''}\delta_{\boldsymbol{k}''\boldsymbol{k}}$.
The thermodynamic limit is taken after discretization, resulting in $[O(\boldsymbol{k})]_{n'n}\delta_{\boldsymbol{k}'\boldsymbol{k}}$.

Three critical conditions regarding the discretization process must be strictly satisfied.
This discretization procedure is valid provided that (i) no additional Dirac delta functions appear in subsequent calculation steps, (ii) the pair of $\Delta_{N}$ functions appears under an integral over their common variable, and (iii) the $\Delta_{N}$ functions remain undifferentiated with respect to the integration variable to be discretized.
The first condition prevents the post-regularization emergence of Dirac delta functions that do not originate from the $\Delta_{N}$ terms.
The second condition forbids discretizing the nonzero surface term, an operation that is mathematically as ill-defined as evaluating the Dirac delta function at a specific discontinuous point.
The last condition establishes the role of the $\Delta_{N}$ function as a regularized representation of the Dirac delta function---which facilitates the transfer of the derivative onto adjacent functions through integration by parts---and ensures that the final physical outcome remains independent of the specific functional form chosen for $\Delta_{N}$.
Specifically, one must refrain from treating the explicitly differentiated form of $\Delta_N$ directly as an ordinary function, as doing so would introduce a dependence on the specific functional form of $\Delta_{N}$.
For example, the evaluation of the intraband velocity, $\lim_{N\rightarrow\infty}N^{-1}\langle \Psi^{N}_{n\boldsymbol{k}}|\hat{\boldsymbol{v}}|\Psi^{N}_{n\boldsymbol{k}}\rangle$, could yield $-\lim_{N\rightarrow\infty}\frac{2}{N}E_n(\boldsymbol{k})i\boldsymbol{\partial_{k''}}\Delta_{N}(\boldsymbol{k}-\boldsymbol{k}'')|_{\boldsymbol{k}''=\boldsymbol{k}}$.

\section{Evaluation results of the position operator}\label{ASection 5}
In this section, we provide the derivation of the general evaluation rule for operator product chains comprising a single position operator and local cell-periodic operators.
For simplicity, we restrict our analysis to a one-dimensional system.
Applying the three rules established previously, we evaluate the product chain $\hat{O}\hat{r}\hat{Q}$ to yield
\begin{equation}\label{General case short}
\begin{split}
    &\lim_{N\rightarrow\infty}\frac{1}{N}\langle \Psi^{N}_{n'k}|\hat{O}\hat{r}\hat{Q}|\Psi^{N}_{nk}\rangle\\
    &=\big[[O(k)][A(k)][Q(k)]\big]_{n'n}\\
    &\quad+\sum_{m}\frac{1}{2}\Big([O(k)]_{n'm}i\partial_{k}[Q(k)]_{mn}\\
    &\quad-(i\partial_{k}[O(k)]_{n'm})[Q(k)]_{mn}\Big).
\end{split}
\end{equation}
The detailed derivation of this result is provided in Appendix~\ref{Appendix A}.
This result can be generalized to three dimensions ($\hat{\boldsymbol{r}}$) by replacing $A$, $k$, and $\partial_{k}$ with $\boldsymbol{A}$, $\boldsymbol{k}$, and $\boldsymbol{\partial_{k}}$, respectively.
Remarkably, this result illustrates that the evaluation of an operator product is not necessarily equivalent to the product of the individual evaluations, $\big[[O(k)][A(k)][Q(k)]\big]_{n'n}$, due to the additional $k$-derivative contributions within the parentheses~\footnote{
When unbounded operators---such as the position operator---are involved in the operator product, the cyclic property of the thermodynamic trace functional may be inapplicable since the operator product does not belong to the trace class.
The sign difference between the $k$-derivatives acting on $[O(k)]$ and $[Q(k)]$ in Eq.~\eqref{General case short} explicitly demonstrates the violation of the cyclic property of the thermodynamic trace functional $\lim_{N\rightarrow\infty}N^{-1}\text{Tr}(\cdot)=\sum_{n}\int\frac{d k}{V_{\text{B.Z.}}}\lim_{N\rightarrow\infty}\frac{1}{N}\langle \Psi^{N}_{nk}|\cdot|\Psi^{N}_{nk}\rangle$.
}.

Equation~\eqref{General case short} directly demonstrates that this evaluation satisfies the requisite physical conditions.
First, setting $\hat{O}=\hat{Q}=1$ recovers the evaluation of the position operator, $\lim_{N\rightarrow\infty}N^{-1}\langle \Psi^{N}_{n'k}|\hat{r}|\Psi^{N}_{nk}\rangle=[A(k)]_{n'n}$, which clearly remains independent of the choice of unit cell.
Second, taking the complex conjugate of Eq.~\eqref{General case short} verifies that the Hermitian conjugate relation for the operator product involving the position operator is strictly preserved; i.e., $\lim_{N\rightarrow\infty}N^{-1}\langle \Psi^{N}_{n'k}|\hat{O}\hat{r}\hat{Q}|\Psi^{N}_{nk}\rangle=\lim_{N\rightarrow\infty}N^{-1}\langle \Psi^{N}_{nk}|\hat{Q}\hat{r}\hat{O}|\Psi^{N}_{n'k}\rangle^{*}$.
Lastly, employing Eq.~\eqref{General case short} confirms that the intraband velocity is correctly evaluated as $\lim_{N\rightarrow\infty}N^{-1}\langle \Psi^{N}_{nk}|\hat{v}|\Psi^{N}_{nk}\rangle=\hbar^{-1}\partial_k E_{n}(k)$, consistent with Eq.~\eqref{Expected position operator}.

Our three rules consistently yield results for the cell-periodic local operator $\hat{O}$ that are identical to those obtained from standard evaluations: $\lim_{N\rightarrow\infty}N^{-1}\langle \Psi^{N}_{n'\boldsymbol{k}}|\hat{O}|\Psi^{N}_{n\boldsymbol{k}}\rangle=\langle \psi_{n'\boldsymbol{k}}|\hat{O}|\psi_{n\boldsymbol{k}}\rangle_{\text{cell}}$.
Notably, in the absence of these rules, critical problems would arise concerning the unit-cell dependence of the evaluations and the proper reproduction of the intraband velocity.
Consequently, our rules, summarized in Table~\ref{Tab: Rules}, provide a robust evaluation framework free from such inconsistencies (see Appendix~\ref{Appendix A} for details).

\section{Gauge dependence of position operator}\label{BSection 2}
Having established the rules for evaluating the position operator and verified their consistency with physical conditions, we now turn our attention to the subtle issue of gauge dependence.
Given that the expectation value of the position operator depends on the choice of origin, its evaluation does not correspond to a directly measurable physical observable.
In other words, the choice of origin is associated with a gauge transformation.
For instance, the gauge choice $\chi_{n}(\boldsymbol{k})=\boldsymbol{k}\cdot\boldsymbol{R}$, where $\boldsymbol{R}$ denotes a lattice vector, for a band-diagonal gauge transformation of the form $|\tilde{\psi}_{n\boldsymbol{k}}\rangle=e^{i\chi_{n}(\boldsymbol{k})}|\psi_{n\boldsymbol{k}}\rangle$, is mathematically equivalent to redefining the real-space origin as $\tilde{\mathcal{O}}=\mathcal{O}-\boldsymbol{R}$:
\begin{equation}
    \langle \boldsymbol{r}|\tilde{\psi}_{n\boldsymbol{k}}\rangle_{\mathcal{O}}=\langle \boldsymbol{r}+\boldsymbol{R}|\psi_{n\boldsymbol{k}}\rangle_{\mathcal{O}}=\langle \boldsymbol{r}|\psi_{n\boldsymbol{k}}\rangle_{\tilde{\mathcal{O}}},
\end{equation}
where the subscript denotes the choice of origin.
This equivalence elucidates the intrinsic link between the origin dependence and the gauge dependence of the position operator.
By extension, composite operators containing the position operator inherit this gauge dependence.
Specifically, under a band-diagonal gauge transformation, the Berry connection and the differentiated operator transform as follows:
\begin{equation}\label{gauge transformation of operators 1}
\begin{split}
    [\tilde{\boldsymbol{A}}(\boldsymbol{k})]_{n'n}&= e^{i(\chi_{n}(\boldsymbol{k})-\chi_{n'}(\boldsymbol{k}))}[\boldsymbol{A}(\boldsymbol{k})]_{n'n}\\
    &\quad-\boldsymbol{\partial_{k}}\chi_{n}(\boldsymbol{k})\delta_{n'n},\\
    i\boldsymbol{\partial_{k}}[\tilde{O}(\boldsymbol{k})]_{n'n}
    &= e^{i(\chi_{n}(\boldsymbol{k})-\chi_{n'}(\boldsymbol{k}))}\Big(i\boldsymbol{\partial_{k}}[O(\boldsymbol{k})]_{n'n}\\
    &\quad + [O(\boldsymbol{k})]_{n'n}\boldsymbol{\partial_{k}}\big(\chi_{n'}(\boldsymbol{k})-\chi_{n}(\boldsymbol{k})\big)\Big),
\end{split}
\end{equation}
and
\begin{equation}\label{gauge transformation of operators 2}
\begin{split}
    i\partial_{k}^{u}[\tilde{A}^{w}(\boldsymbol{k})]_{n'n}
    &= e^{i(\chi_{n}(\boldsymbol{k})-\chi_{n'}(\boldsymbol{k}))}\Big(i\partial_{k}^{u}[A^{w}(\boldsymbol{k})]_{n'n}\\
    &+ [A^{w}(\boldsymbol{k})]_{n'n}\partial_{k}^{u}\big(\chi_{n'}(\boldsymbol{k})-\chi_{n}(\boldsymbol{k})\big)\Big)\\
    &-i\partial_{k}^{u}\partial_{k}^{w}\chi_{n}(\boldsymbol{k})\delta_{n'n},
\end{split}
\end{equation}
where the tildes denote quantities evaluated using the gauge-transformed Bloch states $|\tilde{\psi}_{n\boldsymbol{k}}\rangle$, and $u$ and $w$ are Cartesian indices.
For the class of observables considered here, such as diagonal expectation values and transition probabilities, the overall phase factors $e^{i(\chi_{n}(\boldsymbol{k})-\chi_{n'}(\boldsymbol{k}))}$ cancel out and do not contribute to the gauge dependence.
In these cases, the residual gauge dependence arises primarily through the differentiated gauge functions $\boldsymbol{\partial_{k}}\chi_{n}(\boldsymbol{k})$.
In the following section, we introduce a systematic method to remove this residual gauge dependence.

\section{Gauge filtration}\label{BSection 3}
When the evaluation of a composite operator may exhibit gauge dependence, we subtract its gauge dependence to isolate the underlying physical quantity.
Motivated by the relation of mechanical and canonical momentum in conventional quantum mechanics~\cite{Sakurai2020Book}, we devise a gauge-removal process to filter out this residual dependence following the operator's evaluation, which we term gauge filtration.

\subsection{Steps of gauge filtration}\label{Sec: GF steps}
Gauge filtration is conducted through the following three steps:
\begin{enumerate}
    \item Evaluate the operator using the three rules established in Sec.~\ref{ASection 4} to obtain the expectation value, which may depend on the gauge choice.
    \item Apply the gauge transformation $|\tilde{\psi}_{n\boldsymbol{k}}\rangle = e^{i\chi_{n}(\boldsymbol{k})}|\psi_{n\boldsymbol{k}}\rangle$ for a test function $\chi_{n}(\boldsymbol{k})$ and systematically collect the $\chi_{n}(\boldsymbol{k})$-dependent terms.
    \item Remove the gauge dependence of the evaluated operator by adding a counterterm that exactly cancels the gauge-dependent contributions.
    This counterterm---constructed from the Berry connection and its derivatives---is defined to ensure it does not alter any gauge-invariant physical contributions.
\end{enumerate}

\subsection{Application of gauge filtration}
Building upon the gauge transformation properties discussed in Sec.~\ref{BSection 2}, we apply gauge filtration to the quantities in Eq.~\eqref{gauge transformation of operators 1} as follows:
\begin{equation}\label{Gauge Filtration 1}
\begin{split}
    \big([\boldsymbol{A}(\boldsymbol{k})]_{n'n}\big)^{\text{G.F.}}
    &=[\boldsymbol{A}(\boldsymbol{k})]_{n'n}-\boldsymbol{A}_{n}(\boldsymbol{k})\delta_{n'n},\\
    \big(i\boldsymbol{\partial_{k}}[O(\boldsymbol{k})]_{n'n}\big)^{\text{G.F.}}
    &=i\boldsymbol{\partial_{k}}[O(\boldsymbol{k})]_{n'n}\\
    &+[O(\boldsymbol{k})]_{n'n}\big(\boldsymbol{A}_{n'}(\boldsymbol{k})-\boldsymbol{A}_{n}(\boldsymbol{k})\big),\\
\end{split}
\end{equation}
where the superscript $\text{G.F.}$ denotes the gauge-filtered quantity, and $\boldsymbol{A}_n(\boldsymbol{k})=[\boldsymbol{A}(\boldsymbol{k})]_{nn}$ is the intraband Berry connection.
In these examples, gauge filtration effectively amounts to a direct replacement of $\boldsymbol{\partial_{k}}\chi_{n}(\boldsymbol{k})$ with $\boldsymbol{A}_{n}(\boldsymbol{k})$.
In other words, the physical contribution in the Berry connection matrix is restricted to the band-off-diagonal Berry connection.

However, this simple substitution is insufficient for the differentiated Berry connection in Eq.~\eqref{gauge transformation of operators 2}.
Properly filtering the differentiated Berry connection requires incorporating gauge-invariant quantities---specifically, the Berry curvature and the quantum metric---as follows:
\begin{equation}\label{Gauge Filtration 2}
\begin{split}
    &\big(i\partial_{k}^{u}[A^{w}(\boldsymbol{k})]_{n'n}\big)^{\text{G.F.}}
    =i\partial_{k}^{u}[A^{w}(\boldsymbol{k})]_{n'n}\\
    &+\big(A^{u}_{n'}(\boldsymbol{k})-A^{u}_{n}(\boldsymbol{k})\big)[A^{w}(\boldsymbol{k})]_{n'n}\\
    &-\Big(
    i\partial_{k}^{u}A^{w}_{n}(\boldsymbol{k})
    -\frac{i}{2}\varepsilon^{uw\alpha}\Omega^{\alpha}_{n}(\boldsymbol{k})
    +\frac{1}{2}G^{uw}_{n}(\boldsymbol{k})
    \Big)\delta_{n'n},
\end{split}
\end{equation}
where $\varepsilon^{uw\alpha}$ is the Levi-Civita symbol, $u$, $w$, and $\alpha$ are Cartesian indices, summation over a repeated Cartesian index is implied, $\boldsymbol{\Omega}_{n}(\boldsymbol{k})=\boldsymbol{\partial_{k}}\times \boldsymbol{A}_n(\boldsymbol{k})$ is the Berry curvature of the $n$-th band, and $G_{n}^{uw}(\boldsymbol{k})=\sum_{m\neq n}\big([A^{u}(\boldsymbol{k})]_{nm}[A^{w}(\boldsymbol{k})]_{mn}+[A^{w}(\boldsymbol{k})]_{nm}[A^{u}(\boldsymbol{k})]_{mn}\big)$ is the $n$-th band quantum metric~\cite{Provost1980CommunMathPhys.76.289, Anandan1990PhysRevLett.65.1697, Mera2022PhysRevB.106.165133, Avdoshkin2023PhysRevB.107.245136, Piechon2016PhysRevB.94.134423, Ozawa2021PhysRevB.104.045103, Xu2025ArXiv.11425}.
Here, the second term on the right-hand side is the counterterm for the first-order differentiated gauge function in Eq.~\eqref{gauge transformation of operators 2}, which is obtained by a direct replacement of $\boldsymbol{\partial_{k}}\chi_{n}(\boldsymbol{k})$ with $\boldsymbol{A}_{n}(\boldsymbol{k})$.
In the third term, $i\partial_{k}^{u}A^{w}_{n}(\boldsymbol{k})\delta_{n'n}$ can cancel the second-order differentiated gauge function, but it contains gauge-invariant contributions, namely, the Berry curvature and the quantum metric:
\begin{equation}
\begin{split}
    i\partial_{k}^{u}A^{w}_{n}(\boldsymbol{k})
    =&-\langle u_{n\boldsymbol{k}}|\partial_{k}^{u}\partial_{k}^{w} u_{n\boldsymbol{k}}\rangle
    -A^{u}_{n}(\boldsymbol{k})A^{w}_{n}(\boldsymbol{k})\\
    &-\frac{1}{2}G^{uw}_{n}(\boldsymbol{k})
    +\frac{i}{2}\varepsilon^{uw\alpha}\Omega^{\alpha}_{n}(\boldsymbol{k}).
\end{split}
\end{equation}
Thus, the third term incorporates them to avoid removing gauge-independent terms.

\section{Orbital moment operator}\label{BSection 4}
In this section, we present a consistent evaluation of the orbital moment operator and its gauge-independent results including gauge-filtered evaluation, self-rotation, and local circulation.
Specifically, we identify terms previously omitted or treated only implicitly in certain formulations of the self-rotation of wave packets~\cite{Chang1996PhysRevB.53.7010, Xiao2005PhysRevLett.95.137204} and local circulation in the Wannier representation~\cite{Thonhauser2005PhysRevLett.95.137205, Ceresoli2006PhysRevB.74.024408}.
By incorporating these additional terms and taking the Wannier-function limit of the wave packet, we reconcile these two perspectives through an operator-level evaluation.
This procedure reduces the evaluation to a Brillouin-zone integration of the operator matrix element, thereby rendering their mutual consistency explicit.

Established results concerning the self-rotation of wave packets and local circulation in the Wannier representation can be interpreted as gauge-invariant manifestations of the evaluated orbital moment operator.
From this standpoint, we examine the issues inherent in each approach regarding the process of removing gauge dependence.
For brevity, we henceforth use the terms self-rotation and local circulation to refer to the self-rotation of wave packets and local circulation in the Wannier representation, respectively.

\subsection{Evaluation of the orbital moment operator and its gauge-independent results}
The evaluation of the orbital moment operator---$\hat{\boldsymbol{m}}=(\hat{\boldsymbol{r}}\times\hat{\boldsymbol{v}}-\hat{\boldsymbol{v}}\times\hat{\boldsymbol{r}})/4$---yields the following result for the intraband case:
\begin{equation}\label{Orbital moment equal band Eval}
\begin{split}
    &\lim_{N\rightarrow\infty}\frac{1}{N}\langle \Psi^{N}_{n\boldsymbol{k}}|\hat{\boldsymbol{m}}|\Psi^{N}_{n\boldsymbol{k}}\rangle
    =\frac{1}{2i\hbar}\bigg(
    [\boldsymbol{A}]_{nn}\times i\boldsymbol{\partial_{k}}E_{n}\\
    &\quad+\sum_{m}\big(E_{n}-E_{m}\big)
    [\boldsymbol{A}]_{nm}\times[\boldsymbol{A}]_{mn}\bigg),
\end{split}
\end{equation}
where we omit the explicit $\boldsymbol{k}$ dependence for the band energy and the Berry connection matrix.
As with the position operator, the first term on the right-hand side of Eq.~\eqref{Orbital moment equal band Eval} depends on the choice of gauge.
Performing gauge filtration on the evaluation of the orbital moment operator to extract physically relevant terms is equivalent to removing the gauge dependence from the Berry connection matrix [Eq.~\eqref{Gauge Filtration 1}] because the velocity operator matrix elements are themselves gauge-invariant, $[\boldsymbol{v}(\boldsymbol{k})]_{n'n}=([\boldsymbol{v}(\boldsymbol{k})]_{n'n})^{\text{G.F.}}$.
Consequently, this process leads to
\begin{equation}\label{orbital moment GF}
\begin{split}
    &\big([\boldsymbol{m}(\boldsymbol{k})]_{nn}\big)^{\text{G.F.}}\\
    &=\frac{1}{2i\hbar}\langle\boldsymbol{\partial_{k}} u_{n\boldsymbol{k}}|\times\big(E_{n}(\boldsymbol{k})-\hat{H}(\boldsymbol{k})\big)|\boldsymbol{\partial_{k}}u_{n\boldsymbol{k}}\rangle.
\end{split}
\end{equation}
This result is identical to the Wannier function limit of the self-rotation.
In Sec.~\ref{SR LC gauge-removing perspective}, we demonstrate that the gauge filtration provides an advantage in eliminating gauge dependence compared to the gauge-removal processes specifically contrived to get the self-rotation and the local circulation.

Alongside our gauge-filtered evaluation, two well-known gauge-independent quantities are associated with the orbital moment.
One is the self-rotation~\cite{Chang1996PhysRevB.53.7010, Xiao2005PhysRevLett.95.137204}
\begin{equation}\label{Self-rotation result}
\begin{split}
    &\varepsilon^{uw\alpha}\langle \mathcal{W}|(\hat{r}^{u}-r_{\mathcal{W}}^{u}) \hat{v}^{w}|\mathcal{W}\rangle
    =\int\frac{d\boldsymbol{k}}{V_{\text{B.Z.}}}
    \frac{i}{\hbar}|\mathcal{W}(\boldsymbol{k})|^2\\
    &\times\varepsilon^{uw\alpha}\langle \partial_{k}^{u}u_{n\boldsymbol{k}}| (\hat{H}(\boldsymbol{k})-E_{n}(\boldsymbol{k}))|\partial_{k}^{w} u_{n\boldsymbol{k}}\rangle,
\end{split}
\end{equation}
where $u$, $w$, and $\alpha$ are Cartesian indices, summation over repeated Cartesian indices is implied, $\boldsymbol{r}_{\mathcal{W}}=\langle \mathcal{W}|\hat{\boldsymbol{r}}|\mathcal{W}\rangle=\int\frac{d\boldsymbol{k}}{V_{\text{B.Z.}}}\Big(|\mathcal{W}(\boldsymbol{k})|^2\langle i\boldsymbol{\partial_{k}}u_{n\boldsymbol{k}}|u_{n\boldsymbol{k}}\rangle+\big(-i\boldsymbol{\partial_{k}}\mathcal{W}^{*}(\boldsymbol{k})\big)\mathcal{W}(\boldsymbol{k})\Big)$, $|\mathcal{W}\rangle=\int\frac{d\boldsymbol{k}}{V_{\text{B.Z.}}}\mathcal{W}(\boldsymbol{k})|\psi_{n\boldsymbol{k}}\rangle$, and $\mathcal{W}(\boldsymbol{k})$ is a distribution function defined in the first Brillouin zone, satisfying $\int\frac{d\boldsymbol{k}}{V_{\text{B.Z.}}}|\mathcal{W}(\boldsymbol{k})|^2=1$.

The other is the local circulation~\cite{Thonhauser2005PhysRevLett.95.137205, Ceresoli2006PhysRevB.74.024408},
\begin{equation}\label{Local circulation result}
\begin{split}
    &\lim_{N\rightarrow \infty}\frac{1}{N}\sum_{\boldsymbol{R}}\langle W_{n\boldsymbol{R}}|(\hat{\boldsymbol{r}}-\boldsymbol{r}_{c,n}(\boldsymbol{R}))\times\hat{\boldsymbol{v}}|W_{n\boldsymbol{R}}\rangle\\
    &=\int\frac{d\boldsymbol{k}}{V_{\text{B.Z.}}}
    \frac{i}{\hbar}\langle \boldsymbol{\partial_{k}}u_{n\boldsymbol{k}}|\times \hat{H}(\boldsymbol{k})|\boldsymbol{\partial_{k}} u_{n\boldsymbol{k}}\rangle,
\end{split}
\end{equation}
where $|W_{n\boldsymbol{R}}\rangle$ is the Wannier function defined as $W_{n\boldsymbol{R}}(\boldsymbol{r})=\langle \boldsymbol{r}|W_{n\boldsymbol{R}}\rangle=\int \frac{d\boldsymbol{k}}{V_{\text{B.Z.}}}e^{-i\boldsymbol{k}\cdot\boldsymbol{R}}\psi_{n\boldsymbol{k}}(\boldsymbol{r})$, $\boldsymbol{r}_{c,n}(\boldsymbol{R})=\langle W_{n\boldsymbol{R}}|\hat{\boldsymbol{r}}|W_{n\boldsymbol{R}}\rangle$ is the Wannier center, $n$ is the band index, and $\boldsymbol{R}$ denotes lattice vectors~\cite{Vanderbilt2018Book}.

In Sec.~\ref{missing terms subsection}, we incorporate the missing terms of the self-rotation and local circulation.
We demonstrate that upon including these respective missing terms and taking the Wannier-function limit, Eqs.~\eqref{Self-rotation result} and \eqref{Local circulation result} both identically reduce to the Brillouin-zone integral of Eq.~\eqref{Orbital moment equal band Eval}, resulting in:
\begin{equation}\label{LC SR correct results}
\begin{split}
    &\lim_{N\rightarrow \infty}\frac{1}{N}\int \frac{d\boldsymbol{k}}{V_{\text{B.Z.}}}\langle \Psi^{N}_{n\boldsymbol{k}}|2\hat{\boldsymbol{m}}|\Psi^{N}_{n\boldsymbol{k}}\rangle\\
    &=
    \frac{1}{i\hbar}\int \frac{d\boldsymbol{k}}{V_{\text{B.Z.}}}\bigg(
    [\boldsymbol{A}]_{nn}\times i\boldsymbol{\partial_{k}}E_{n}\\
    &\quad+\sum_{m}\big(E_{n}-E_{m}\big)
    [\boldsymbol{A}]_{nm}\times[\boldsymbol{A}]_{mn}\bigg),
\end{split}
\end{equation}
where the intraband evaluations of $2\hat{\boldsymbol{m}}(\boldsymbol{k})$ and $\hat{\boldsymbol{r}}\times\hat{\boldsymbol{v}}$ are identical despite the fact that $2\hat{\boldsymbol{m}}(\boldsymbol{k})\neq \hat{\boldsymbol{r}}\times\hat{\boldsymbol{v}}$.

\subsection{Missing terms in the self-rotation and local circulation}\label{missing terms subsection}
To see the missing terms explicitly, we first note that the missing term for the orbital moment in the result [Eq.~\eqref{Self-rotation result}] from the self-rotation~\cite{Chang1996PhysRevB.53.7010} is
\begin{equation}\label{SR error 1}
\begin{split}
    &-\int\frac{d\boldsymbol{k}}{V_{\text{B.Z.}}}\boldsymbol{v}_{n}(\boldsymbol{k})\times \bigg(
    |\mathcal{W}(\boldsymbol{k})|^2\langle i\boldsymbol{\partial_{k}}u_{n\boldsymbol{k}}|u_{n\boldsymbol{k}}\rangle\\
    &+\big(-i\boldsymbol{\partial_{k}}\mathcal{W}^{*}(\boldsymbol{k})\big)\mathcal{W}(\boldsymbol{k})
    -|\mathcal{W}(\boldsymbol{k})|^2\boldsymbol{r}_{\mathcal{W}}
    \bigg),
\end{split}
\end{equation}
which was implicitly obtained in Ref.~\cite{Chang1996PhysRevB.53.7010} along with Eq.~\eqref{Self-rotation result}, but the authors assumed it to be zero.
This assumption does not hold universally under rigorous evaluation because the correlation between the velocity and position of the wave packet captured by this missing term does not have to be zero without any restriction.
Incorporating the nonzero contribution from Eq.~\eqref{SR error 1} into Eq.~\eqref{Self-rotation result} recovers Eq.~\eqref{LC SR correct results} in the limit $\mathcal{W}(\boldsymbol{k})\rightarrow e^{i\boldsymbol{k}\cdot \boldsymbol{R}}$~(see Appendix~\ref{Missing term in SR}).

Next, we direct our attention to the missing terms in the local circulation result of Eq.~\eqref{Local circulation result}.
We evaluate the the left-hand side of Eq.~\eqref{Local circulation result} in the Wannier representation by replacing the Wannier states with $\Delta_N$-regularized Wannier states $|\mathfrak{W}^{N}_{n\boldsymbol{R}}\rangle=\int \frac{d\boldsymbol{k}}{V_{\text{B.Z.}}}e^{-i\boldsymbol{k}\cdot\boldsymbol{R}}|\Psi^{N}_{n\boldsymbol{k}}\rangle$.
In this process, we apply the proposed three rules for evaluation in the Bloch representation to the $\Delta_N$-regularized Bloch states within the $\Delta_N$-regularized Wannier states~\footnote{
Upon taking the thermodynamic limit, the Dirac delta function $\delta(\boldsymbol{k}'-\boldsymbol{k})$ emerges, provided that $\lim_{N\rightarrow\infty} N \delta_{\boldsymbol{k}'\boldsymbol{k}}/V_{\text{B.Z.}}$ exists.
}.
In this approach, the local circulation result $\lim_{N\rightarrow\infty}N^{-1}\sum_{\boldsymbol{R}}\langle \mathfrak{W}^{N}_{n\boldsymbol{R}}|2\hat{\boldsymbol{m}}|\mathfrak{W}^{N}_{n\boldsymbol{R}}\rangle$ is identical to Eq.~\eqref{LC SR correct results}, i.e., there is no missing term.

To clarify the missing term of the local circulation, we analyze the difference between the right-hand sides of Eq.~\eqref{Local circulation result} and Eq.~\eqref{LC SR correct results}, $\hbar^{-1}\int \frac{d\boldsymbol{k}}{V_{\text{B.Z.}}} \boldsymbol{\partial_{k}}\times\big(E_{n}(\boldsymbol{k})[\boldsymbol{A}(\boldsymbol{k})]_{nn}\big)$.
The issue of this omitted surface term depends on the system.
Specifically, while this missing term is irrelevant for a trivial insulator because it vanishes, it becomes a nonzero missing component of the local circulation for a Chern insulator.
In the latter case, this term carries gauge dependence because the surface term depends on the gauge choice (see Appendix~\ref{Missing term in LC}).

The omission of this surface term seems to have originated from applying the following evaluation, which induces an implicit integration by parts that removes the surface term:
\begin{equation}\label{r eval in Wannier}
    \langle W_{n'\boldsymbol{R}'}|\hat{\boldsymbol{r}}|W_{n\boldsymbol{R}}\rangle=[\boldsymbol{A}(\boldsymbol{R}'-\boldsymbol{R})]_{n'n}+\delta_{n'n}\delta_{\boldsymbol{R}'\boldsymbol{R}}\boldsymbol{R}.
\end{equation}
For instance, evaluating operators $\langle W_{n\boldsymbol{R}'}|\{\hat{\boldsymbol{r}},\hat{O}\}_{-}|W_{n\boldsymbol{R}}\rangle$ such as the velocity operator using Eq.~\eqref{r eval in Wannier} produces the $\int \frac{d\boldsymbol{k}}{V_{\text{B.Z.}}}e^{i\boldsymbol{k}\cdot(\boldsymbol{R}'-\boldsymbol{R})}(\boldsymbol{R}'-\boldsymbol{R})[O(\boldsymbol{k})]_{nn}$ term, whereas the evaluation using $\Delta_N$-regularized Wannier states yields $\int \frac{d\boldsymbol{k}}{V_{\text{B.Z.}}}e^{i\boldsymbol{k}\cdot(\boldsymbol{R}'-\boldsymbol{R})}i\boldsymbol{\partial_{k}}[O(\boldsymbol{k})]_{nn}$.
For a cell‑periodic matrix $[O(\boldsymbol{k})]$, which is surface-term-free over the Brillouin zone, the two results are identical.
If, however, $[O(\boldsymbol{k})]$ is not surface-term-free over the Brillouin zone, then the two results differ due to the non-vanishing surface term:
\begin{equation}
\begin{split}
    &\int \frac{d\boldsymbol{k}}{V_{\text{B.Z.}}}e^{i\boldsymbol{k}\cdot(\boldsymbol{R}'-\boldsymbol{R})}(\boldsymbol{R}'-\boldsymbol{R})[O(\boldsymbol{k})]_{nn}\\
    &\neq\int \frac{d\boldsymbol{k}}{V_{\text{B.Z.}}}e^{i\boldsymbol{k}\cdot(\boldsymbol{R}'-\boldsymbol{R})}i\boldsymbol{\partial_{k}}[O(\boldsymbol{k})]_{nn}.
\end{split}
\end{equation}
This issue emerges if the $\delta_{n'n}\delta_{\boldsymbol{R}'\boldsymbol{R}}\boldsymbol{R}$ term involves an operator that is not surface-term-free over the Brillouin zone, such as the Berry connection $[\boldsymbol{A}(\boldsymbol{k})]_{nn}$ in a Chern insulator.
This discrepancy becomes evident in the evaluation of $\hat{\boldsymbol{r}}\times \hat{\boldsymbol{r}}$.
The evaluation using Eq.~\eqref{r eval in Wannier} yields
\begin{equation}\label{wrong eval in Wannier with r}
\begin{split}
    &\lim_{N\rightarrow \infty}\frac{1}{N}\sum_{\tilde{\boldsymbol{R}}}
    \langle W_{n(\boldsymbol{R}'+\tilde{\boldsymbol{R}})}|
    \hat{\boldsymbol{r}}\times \hat{\boldsymbol{r}}
    |W_{n(\boldsymbol{R}+\tilde{\boldsymbol{R}})}\rangle\big|_{\boldsymbol{R}'=\boldsymbol{R}} \neq 0.
\end{split}
\end{equation}
In contrast, using the $\Delta_N$-regularized Wannier states, we obtain 
\begin{equation}\label{correct eval in Wannier with r}
\begin{split}
    &\lim_{N\rightarrow \infty}\frac{1}{N}\sum_{\tilde{\boldsymbol{R}}}
    \langle \mathfrak{W}^{N}_{n(\boldsymbol{R}'+\tilde{\boldsymbol{R}})}|
    \hat{\boldsymbol{r}}\times \hat{\boldsymbol{r}}
    |\mathfrak{W}^{N}_{n(\boldsymbol{R}+\tilde{\boldsymbol{R}})}\rangle\big|_{\boldsymbol{R}'=\boldsymbol{R}} = 0,
\end{split}
\end{equation}
which is identically the correct result obtained by inserting the identity operator $\int d\boldsymbol{r}|\boldsymbol{r}\rangle\langle\boldsymbol{r}|$ between the two position operators.

To summarize, once the respective missing terms are properly included, the self-rotation and the local circulation yield the same Brillouin-zone integral for the orbital moment in the Wannier-function limit, and thus become equivalent descriptions of the same physical quantity~\footnote{
Because the self-rotation, upon incorporating the missing terms in the Wannier-function limit, remains independent of $\boldsymbol{R}$, applying a normalized summation over the lattice vectors---as performed in the evaluation of the local circulation---does not alter the result.
}.
Since these results manifest gauge dependence in the context of Chern insulators, an auxiliary procedure is required to eliminate this dependency and extract a physically observable quantity.
In this sense, the self-rotation in Eq.~\eqref{Self-rotation result} and local circulation in Eq.~\eqref{Local circulation result} may be interpreted as the gauge-invariant manifestations of Eq.~\eqref{LC SR correct results}.
We will elaborate upon their gauge-removal processes and provide a comparative analysis with the gauge filtration in Sec.~\ref{SR LC gauge-removing perspective}.

\subsection{Self-rotation and local circulation from a gauge-removal perspective}\label{SR LC gauge-removing perspective}
Since the self-rotation and local circulation---including missing terms---are gauge-dependent (see Eq.~\eqref{LC SR correct results}), a gauge-removal process is necessary to obtain physically meaningful results representing these quantities.
From a gauge-removal perspective, the self-rotation in Eq.~\eqref{Self-rotation result} and local circulation in Eq.~\eqref{Local circulation result} can be interpreted as gauge-invariant results of Eq.~\eqref{LC SR correct results} derived through distinct gauge-removal procedures.
Here, we construct the specific gauge-removing processes that reproduce the self-rotation in Eq.~\eqref{Self-rotation result} and the local circulation in Eq.~\eqref{Local circulation result}, and demonstrate their shortcomings compared to gauge filtration.

For the self-rotation, we can remove the gauge dependence by applying gauge filtration or by replacing the position operator with a gauge-independent position operator $\hat{\boldsymbol{\mathcal{R}}}=\hat{\boldsymbol{r}}-\sum_{n}\int \frac{d\boldsymbol{k}}{V_{\text{B.Z.}}} [\boldsymbol{A}(\boldsymbol{k})]_{nn}|\psi_{n\boldsymbol{k}}\rangle\langle\psi_{n\boldsymbol{k}}|$~\cite{Ibanez-Azpiroz2022SciPostPhys.12.070, Go2024PhysRevB.109.174435, Xu2025ArXiv.11425}.
We refer to the latter as the self-rotation gauge-removal scheme to distinguish it from gauge filtration.
For the local circulation, the gauge dependence resides in the surface term omitted from Eq.~\eqref{Local circulation result}.
Thus, we define the removal of this gauge-dependent surface term as the local circulation gauge-removal scheme.

The self-rotation and local circulation gauge-removal schemes successfully remove the gauge dependence when evaluating the orbital moment itself. 
However, these schemes cannot fully eliminate the gauge dependence of the quantities derived from the orbital moment such as $\{\hat{O},\hat{\boldsymbol{m}}\}_{+}$; a prime example is the $x$-directional orbital moment current $\hat{j}^{x}_{\boldsymbol{m}}=(\hat{\boldsymbol{m}}\hat{v}^{x}+\hat{v}^{x}\hat{\boldsymbol{m}})/2$.
In the self-rotation gauge-removal scheme, the orbital moment is substituted by $\hat{\boldsymbol{m}}^{\text{S.R.}}=(\hat{\boldsymbol{\mathcal{R}}}\times \hat{\boldsymbol{v}}-\hat{\boldsymbol{v}}\times \hat{\boldsymbol{\mathcal{R}}})/4$, which corresponds to evaluating $\lim_{N\rightarrow\infty}N^{-1}\sum_{\boldsymbol{R}}\langle \mathfrak{W}^{N}_{n\boldsymbol{R}}|\{\hat{O},\hat{\boldsymbol{m}}^{\text{S.R.}}\}_{+}|\mathfrak{W}^{N}_{n\boldsymbol{R}}\rangle$.
For $\hat{O}=\hat{v}^{x}/2$, this yields:
\begin{equation}
\begin{split}
    &\lim_{N\rightarrow\infty}\frac{1}{N}\sum_{\boldsymbol{R}}\langle \mathfrak{W}^{N}_{n\boldsymbol{R}}|\{\hat{v}^{x}/2,\hat{\boldsymbol{m}}^{\text{S.R.}}\}_{+}|\mathfrak{W}^{N}_{n\boldsymbol{R}}\rangle\\
    &=\frac{1}{2}\int \frac{d\boldsymbol{k}}{V_{\text{B.Z.}}}\Big(
    \big[\{[v^{x}(\boldsymbol{k})],([\boldsymbol{m}(\boldsymbol{k})])^{\text{G.F.}}\}_{+}\big]_{nn}\\
    &\quad-\frac{1}{4}\sum_{m}\big(i\boldsymbol{\partial_{k}}[v^{x}]_{nm}\times[\boldsymbol{v}(\boldsymbol{k})]_{mn}\\
    &\quad+[\boldsymbol{v}(\boldsymbol{k})]_{nm}\times i\boldsymbol{\partial_k}[v^{x}]_{mn}\big)
    \Big).
\end{split}
\end{equation}
Here, although the gauge dependence of the orbital moment evaluation is eliminated (i.e., $[\boldsymbol{m}(\boldsymbol{k})]$ is replaced by $([\boldsymbol{m}(\boldsymbol{k})])^{\text{G.F.}}$ in $[\{[v^{x}(\boldsymbol{k})],[\boldsymbol{m}(\boldsymbol{k})]\}_{+}]_{nn}$), the Berry connection contribution arising from $i\boldsymbol{\partial_k}[v^{x}]_{nm}=[i\boldsymbol{\partial_{k}}v^{x}(\boldsymbol{k})]_{nm}-\big[\big\{[\boldsymbol{A}(\boldsymbol{k})],[v^{x}(\boldsymbol{k})]\big\}_{-}\big]_{nm}$ remains uncancelled.

In the local circulation gauge-removal scheme, we remove the gauge dependence by neglecting the gauge-dependent surface term in the evaluation of $\{\hat{O},\hat{\boldsymbol{m}}\}_{+}$, which takes the form:
\begin{equation}\label{LC GR scheme gauge dep result}
\begin{split}
    &\frac{1}{2}\sum_{m}\int \frac{d\boldsymbol{k}}{V_{\text{B.Z.}}}
    \Big([\boldsymbol{A}(\boldsymbol{k})]_{nn}\times[\boldsymbol{v}(\boldsymbol{k})]_{nm}[O(\boldsymbol{k})]_{mn}
    \\
    &\quad+[O(\boldsymbol{k})]_{nm}[\boldsymbol{A}(\boldsymbol{k})]_{mm}\times [\boldsymbol{v}(\boldsymbol{k})]_{mn}\Big).
\end{split}
\end{equation}
For instance, when evaluating the $x$-directional orbital moment current, the $m=n$ contribution in Eq.~\eqref{LC GR scheme gauge dep result} becomes:
\begin{equation}\label{LC GR scheme gauge dep terms of jxm}
\begin{split}
    &\frac{1}{2i\hbar}\int \frac{d\boldsymbol{k}}{V_{\text{B.Z.}}}[v^{x}(\boldsymbol{k})]_{nn}[\boldsymbol{A}(\boldsymbol{k})]_{nn}\times i\boldsymbol{\partial_{k}}E_{n}(\boldsymbol{k})\\
    &=\frac{1}{2i\hbar}\int \frac{d\boldsymbol{k}}{V_{\text{B.Z.}}}\Big(
    E_{n}(\boldsymbol{k})\big(i\boldsymbol{\partial_{k}}[v^{x}(\boldsymbol{k})]_{nn}\big)\times[\boldsymbol{A}]_{nn}\\
    &+E_{n}(\boldsymbol{k})[v^{x}(\boldsymbol{k})]_{nn}i\boldsymbol{\partial_{k}}\times[\boldsymbol{A}]_{nn}\\
    &-i\boldsymbol{\partial_{k}}\times\big(E_{n}(\boldsymbol{k})[v^{x}(\boldsymbol{k})]_{nn}[\boldsymbol{A}]_{nn}\big)
    \Big).
\end{split}
\end{equation}
The local circulation gauge-removal scheme neglects the last term in Eq.~\eqref{LC GR scheme gauge dep terms of jxm}.
However, the term $\frac{1}{2i\hbar}\int \frac{d\boldsymbol{k}}{V_{\text{B.Z.}}}E_{n}(\boldsymbol{k})\big(i\boldsymbol{\partial_{k}}[v^{x}(\boldsymbol{k})]_{nn}\big)\times[\boldsymbol{A}]_{nn}$ remains. Thus, this scheme fails to completely eliminate the gauge dependence.
Moreover, additional Berry connection terms arise from the momentum derivatives of the operator evaluations (see Appendix~\ref{LC SR Gauge filtration}).

We observe that when a gauge-removal scheme is tailored to eliminate the gauge dependence of a specific instance, e.g., the self-rotation gauge-removing scheme and local circulation gauge-removing scheme to eliminate that of the orbital moment operator, it typically does not extend to more general cases.
This limitation arises because derivatives appearing in the evaluation of composite operators introduce additional gauge-dependent terms that are not accounted for in the original construction.
In contrast, within the class of composite operators analyzed in this work, gauge filtration provides a systematic methodology to remove all residual band-diagonal gauge dependence stemming from the position operator.
Therefore, the gauge filtration yields gauge-invariant results even in situations where ad hoc prescriptions---such as the self-rotation and local-circulation schemes---may prove inadequate.

\section{Discussion}\label{Sect: Discussion}
While the $\Delta_N$-regularization scheme is applicable to a broad class of observables, it is not universally applicable to all possible operators.
Its primary limitation stems from the third rule, which cannot be straightforwardly applied to arbitrary composite operators containing multiple position operators.
During the evaluation of a composite operator containing multiple position operators, the order of the derivatives acting on the $\Delta_{N}$ function can accumulate as derivatives are transferred via integration by parts, yielding expressions such as:
\begin{equation}\label{Eq: MDOD ex}
\begin{split}
    &\lim_{N\rightarrow\infty}\frac{V_{\text{B.Z.}}}{N}\int dk'[O(k')]_{n'n}\\
    &\quad\times\Delta_{N}(k''-k')(i\partial_{k'})^a\Delta_{N}(k-k').
\end{split}
\end{equation}
Within the $\Delta_{N}$-regularization framework, evaluating such instances becomes mathematically intractable if the order of derivatives acting on the $\Delta_{N}$ function exceeds one.
A prime example is the evaluation of $\lim_{N\rightarrow\infty}N^{-1}\langle \Psi^{N}_{nk}|\hat{r}^{2}|\Psi^{N}_{nk}\rangle$.
In these cases, the systematic removal of derivatives acting on the $\Delta_{N}$ function is precluded, rendering the third rule inapplicable.

Consequently, the three rules reliably apply only to operators where, for all non-vanishing terms in the evaluation, the maximum derivative order $a$ in Eq.~\eqref{Eq: MDOD ex} is restricted to 0 or 1, with the additional requirement that $[O(k')]_{n'n}$ must be surface-term-free over the Brillouin zone, specifically when $a=1$.
This valid regime encompasses all cell-periodic local operators, the position operator, the velocity operator, the orbital moment operator $\hat{\boldsymbol{m}}$, and the $x$-directional current of the $z$-component of the orbital moment, $\hat{j}^{x}_{m^{z}}=\{\hat{v}^{x},\hat{m}^{z}\}_{+}/2$.
In a general $d$-dimensional system, this limitation criterion is determined by the maximum derivative order observed across all Cartesian directions.
Specifically, the criterion requires $\max_{i}(a_i) \leq 1$, where $a_i$ denotes the maximum $i$-directional derivative order acting on the one-dimensional distribution $\Delta_{N}(k^{i})$ within the factored product $\Delta_{N}(\boldsymbol{k})= \prod_i \Delta_{N}(k^{i})$.

\section{Conclusion}\label{ASection 7}
In this work, we have proposed a methodology for evaluating composite operators involving the position operator---a fundamental quantity broadly used in condensed matter physics, yet whose calculation remains ill-defined in the Bloch representation.
We have established three rules that provide consistent results for handling the position operator within the Bloch representation.
These rules successfully satisfy physical conditions: the evaluation of the position operator remains independent of the choice of unit cell, the Hermitian conjugate relation is preserved, and the intraband velocities are correctly recovered.
Furthermore, we have introduced a gauge-removal scheme, termed gauge filtration, to address the inherent gauge dependence of the position operator, which otherwise introduces spurious gauge-dependent terms into the evaluation of composite operators.
By applying this filtration, we have demonstrated that it is possible to isolate physically measurable quantities, such as the orbital moment, from the evaluated operator expressions.

Using the proposed three rules, we have performed a consistent evaluation of the orbital moment operator.
Our analysis clarifies how the Brillouin-zone integral of the orbital moment operator reproduces both the self-rotation of wave packets~\cite{Chang1996PhysRevB.53.7010, Xiao2005PhysRevLett.95.137204, Xiao2010RevModPhys.82.1959} in the Wannier-function limit and the local circulation in the Wannier representation~\cite{Thonhauser2005PhysRevLett.95.137205, Ceresoli2006PhysRevB.74.024408}.
We have achieved this by explicitly identifying---at the operator-evaluation level---additional contributions previously omitted or left implicit in those formulations and incorporating them into a unified expression.
To elucidate the origin of the missing terms in the local circulation, we have analyzed the complications arising from the naive substitution of the position operator within the Wannier representation, as given in Eq.~\eqref{r eval in Wannier}.
Our investigation revealed that such issues---particularly in Chern insulators---arise because the conventional substitution implicitly neglects the surface term.
Finally, we have applied the $\Delta_N$-regularization to the Wannier representation required for the correct evaluation of the operator, thereby resolving these long-standing issues and achieving accurate results.

The motivation for this study originated from the current lack of a consistent evaluation framework for the position operator in infinitely periodic systems.
We anticipate that our proposed rules will be broadly applicable to various physical operators containing the position operator, including orbital moment currents~\cite{Bhowal2021PhysRevB.103.195309, An2025PhysRevB.111.104436, Go2018PhysRevLett.121.086602, Go2024NanoLett.24.5968} and perturbations involving the position operator in infinitely periodic systems~\cite{Luttinger1964PhysRev.135.A1505, Adams1957PhysRev.107.698, Blount1962PhysRev.126.1636, Karplus1954PhysRev.95.1154, Zak1968PhysRev.168.686, Zak1969PhysRev.177.1151, Hasegawa1969PhysRev.177.1392, Kohn1959PhysRev.115.1460, Adams1959JPhysChemSolids.10.286, Roth1962JPhysChemSolids.23.433,Thouless1982PhysRevLett.49.405}.
Furthermore, this framework can be extended to evaluate unbounded operators in other delocalized periodic bases---such as photonic crystals~\cite{deSterke1988PhysRevA.38.5149, Feng2024JOptSocAmB.41.1471}, magnonic crystals~\cite{Shindou2013PhysRevB.87.174427, Rychly2015LowTempPhys.41.745, Mieszczak2022SciRep.12.11335}, and lattice QCD~\cite{Friedberg1994JMathPhys.35.5600, Friedberg1995PhysRevD.52.4053}.
Nevertheless, we acknowledge a specific limitation of the proposed rules: they are currently restricted to operators where the derivative order of the distribution $\Delta_{N}$ does not exceed one in any direction. 
Properly handling cases involving higher-order distributions requires further mathematical refinement.
We have restricted the system to be nondegenerate in this work.
For degenerate multiband systems, a non-Abelian Berry connection is expected in the evaluation of the position operator.
These limitations and system restrictions are left as an open avenue for future investigation beyond the scope of the current work~\cite{Song2024ArXiv.02519, Juric2022Universe.8.129, Vanderbilt2018Book}.

\section{Acknowledgment}
We thank Seunghun Lee and Geonsu Park for the helpful discussions and comments.
D.A. and S.K.K. were supported by Samsung Science and Technology Foundation (SSTF-BA2202-04), Brain Pool Plus Program through the National Research Foundation of Korea funded by the Ministry of Science and ICT (2020H1D3A2A03099291), and National Research Foundation of Korea(NRF) grant funded by the Korea government(MSIT) (RS-2026-25470048).
J.J. was supported by JSPS KAKENHI Grant Number JP25H01397.

\bibliographystyle{apsrev4-2}
\bibliography{Reference}
\clearpage
\newpage
\begin{appendix}
\begin{widetext}
\renewcommand\thefigure{\thesection\arabic{figure}}

\section{Rationale behind the rules}\label{Appendix A}
This section elucidates the justifications for the evaluation rules established in Sec.~\ref{ASection 4}, which were formulated to satisfy the physical conditions presented in Sec.~\ref{ASection 1}.
Specifically, the evaluation of a cell-periodic local operator must yield identical results regardless of whether one employs the proposed three rules or the standard approach utilizing cell-restricted Bloch states.
Consistent with the assumptions presented in Sec.~\ref{ASection 2}, our analysis focuses on spinless, non-degenerate, infinitely periodic systems, restricting the operators under consideration to either the position operator or cell-periodic local operators.

The first subsection demonstrates that applying the three rules consistently produces results for the cell-periodic local operator $\hat{O}$ that are identical to those obtained from standard evaluations $\langle \psi_{n'\boldsymbol{k}}|\hat{O}|\psi_{n\boldsymbol{k}}\rangle_{\text{cell}}$, i.e., $\lim_{N\rightarrow\infty}N^{-1}\langle \Psi^{N}_{n'\boldsymbol{k}}|\hat{O}|\Psi^{N}_{n\boldsymbol{k}}\rangle=[O(\boldsymbol{k})]_{n'n}$, and clarifies the complexities inherent in evaluating the position operator compared with the standard approach.
The second subsection applies these rules to the general case involving a product chain of operators that contains a single position operator, $\hat{O}\hat{\boldsymbol{r}}\hat{Q}$.

\subsection{Local operator versus position operator}
We first establish that the proposed three rules facilitate a consistent evaluation of a cell-periodic local operator $\hat{O}$, while ensuring that the evaluation of the position operator $\hat{\boldsymbol{r}}$ remains independent of the choice of unit cell.
The evaluation is conducted in the Bloch representation, utilizing the Bloch state expressed as $\psi_{n\boldsymbol{k}}(\boldsymbol{r})=\langle \boldsymbol{r}|\psi_{n\boldsymbol{k}}\rangle=e^{i\boldsymbol{k}\cdot\boldsymbol{r}}u_{n\boldsymbol{k}}(\boldsymbol{r})=\langle \boldsymbol{r}|e^{i\boldsymbol{k}\cdot\hat{\boldsymbol{r}}}|u_{n\boldsymbol{k}}\rangle$, where $n$ is the band index.

\subsubsection{Consistency check}
Conventionally, a cell-periodic local operator $\hat{O}$ is evaluated within a unit cell, where the cell-restricted Bloch states are normalized.
Specifically, when the bra and ket states have the same wave vector $\boldsymbol{k}$, the evaluation of a cell-periodic local operator is given by $\langle \psi_{n'\boldsymbol{k}}|\hat{O}|\psi_{n\boldsymbol{k}}\rangle_{\text{cell}}=\int_{\text{cell}} d\boldsymbol{r} u_{n'\boldsymbol{k}}^*(\boldsymbol{r}) O(\boldsymbol{r}) u_{n\boldsymbol{k}}(\boldsymbol{r})$.
When they have different $\boldsymbol{k}$ values, the matrix element vanishes.
These evaluations follow from the cell periodicity $O(\boldsymbol{r}+\boldsymbol{R})=O(\boldsymbol{r})$ and the lattice sum $\sum_{\boldsymbol{R}}e^{i(\boldsymbol{k}-\boldsymbol{k}')\cdot\boldsymbol{R}}=N\delta_{\boldsymbol{k}\boldsymbol{k}'}$ applied to $\lim_{N\rightarrow\infty}N^{-1}\langle \psi_{n'\boldsymbol{k}}|\hat{O}|\psi_{n\boldsymbol{k}}\rangle
=\lim_{N\rightarrow\infty}N^{-1}\int d\boldsymbol{r}e^{i(\boldsymbol{k}-\boldsymbol{k}')\cdot \boldsymbol{r}} u_{n'\boldsymbol{k}'}^*(\boldsymbol{r}) O(\boldsymbol{r}) u_{n\boldsymbol{k}}(\boldsymbol{r})$.
Here, the continuous-$\boldsymbol{k}$ result is obtained as the continuum limit of the discrete-$\boldsymbol{k}$ evaluation, which is useful for avoiding explicit $N$ factors and Dirac delta functions, e.g., circumventing terms like $V_{\text{B.Z.}}\delta(\boldsymbol{k}-\boldsymbol{k}')/N$.

We can verify the consistency of applying the three rules to evaluate an operator $\hat{O}$ as follows:
\begin{equation}\label{O eval rule 2 explain}
\begin{split}
    &\lim_{N\rightarrow\infty}\frac{1}{N}\langle \Psi^{N}_{n'\boldsymbol{k}'}|\hat{O}|\Psi^{N}_{n\boldsymbol{k}}\rangle
    =\lim_{N\rightarrow\infty}\frac{1}{N}\iint d\tilde{\boldsymbol{k}}'d\tilde{\boldsymbol{k}}\Delta_{N}(\boldsymbol{k}'-\tilde{\boldsymbol{k}}')\Delta_{N}(\boldsymbol{k}-\tilde{\boldsymbol{k}})
    \int d\boldsymbol{r}e^{i(\tilde{\boldsymbol{k}}-\tilde{\boldsymbol{k}}')\cdot \boldsymbol{r}} u_{n'\tilde{\boldsymbol{k}}'}^*(\boldsymbol{r}) O(\boldsymbol{r}) u_{n\tilde{\boldsymbol{k}}}(\boldsymbol{r})\\
    &=\lim_{N\rightarrow\infty}\frac{V_{\text{B.Z.}}}{N}\iint d\tilde{\boldsymbol{k}}'d\tilde{\boldsymbol{k}}\Delta_{N}(\boldsymbol{k}'-\tilde{\boldsymbol{k}}')\Delta_{N}(\boldsymbol{k}-\tilde{\boldsymbol{k}})
    \delta(\tilde{\boldsymbol{k}}-\tilde{\boldsymbol{k}}')\int_{\text{cell}} d\boldsymbol{r}e^{i(\tilde{\boldsymbol{k}}-\tilde{\boldsymbol{k}}')\cdot \boldsymbol{r}} u_{n'\tilde{\boldsymbol{k}}'}^*(\boldsymbol{r}) O(\boldsymbol{r}) u_{n\tilde{\boldsymbol{k}}}(\boldsymbol{r})\\
    &=\lim_{N\rightarrow\infty}\frac{V_{\text{B.Z.}}}{N}\int d\tilde{\boldsymbol{k}}\Delta_{N}(\boldsymbol{k}'-\tilde{\boldsymbol{k}})\Delta_{N}(\boldsymbol{k}-\tilde{\boldsymbol{k}})
    \int_{\text{cell}} d\boldsymbol{r} u_{n'\tilde{\boldsymbol{k}}}^*(\boldsymbol{r}) O(\boldsymbol{r}) u_{n\tilde{\boldsymbol{k}}}(\boldsymbol{r})\\
    &=\lim_{N\rightarrow\infty}
    \sum_{\tilde{\boldsymbol{k}}}\delta_{\boldsymbol{k}'\tilde{\boldsymbol{k}}}\delta_{\tilde{\boldsymbol{k}}\boldsymbol{k}}
    \int_{\text{cell}} d\boldsymbol{r} u_{n'\tilde{\boldsymbol{k}}}^*(\boldsymbol{r}) O(\boldsymbol{r}) u_{n\tilde{\boldsymbol{k}}}(\boldsymbol{r})
    =\delta_{\boldsymbol{k}\boldsymbol{k}'}
    \int_{\text{cell}} d\boldsymbol{r} u_{n'\boldsymbol{k}}^*(\boldsymbol{r}) O(\boldsymbol{r}) u_{n\boldsymbol{k}}(\boldsymbol{r})
    =\delta_{\boldsymbol{k}\boldsymbol{k}'}[O(\boldsymbol{k})]_{n'n}.
\end{split}
\end{equation}
Thus, the three rules yield identical results for any cell-periodic local operator.

\subsubsection{Evaluation of the position operator}
Since the position operator is not cell-periodic, we must use the extended Bloch states, rather than the cell-restricted Bloch states, to evaluate its matrix elements.
In this case, the direct evaluation at the same $\boldsymbol{k}$ using exact Bloch states depends on the choice of unit cell due to the term $\int_\text{cell} d\boldsymbol{r} u_{n\boldsymbol{k}}^*(\boldsymbol{r})u_{n\boldsymbol{k}}(\boldsymbol{r})\boldsymbol{r}$ within the expression $\lim_{N\rightarrow\infty}N^{-1}\langle \psi_{n\boldsymbol{k}}|\hat{\boldsymbol{r}}|\psi_{n\boldsymbol{k}}\rangle =\int_\text{cell} d\boldsymbol{r} u_{n\boldsymbol{k}}^*(\boldsymbol{r})u_{n\boldsymbol{k}}(\boldsymbol{r})\boldsymbol{r}+\bar{\boldsymbol{R}}$, where $\bar{\boldsymbol{R}}$ is defined such that the set $\{\boldsymbol{R}-\bar{\boldsymbol{R}}\mid \boldsymbol{R}\in \text{the set of all lattice vectors} \}$ is invariant under spatial inversion.
Moreover, the intraband velocity cannot be derived from the time derivative of this expression; i.e., $\lim_{N\rightarrow\infty}N^{-1}\frac{d}{dt}\langle \psi_{n\boldsymbol{k}}|\hat{\boldsymbol{r}}|\psi_{n\boldsymbol{k}}\rangle=\int_\text{cell} d\boldsymbol{r}\frac{d}{dt}\big( u_{n\boldsymbol{k}}^*(\boldsymbol{r})u_{n\boldsymbol{k}}(\boldsymbol{r})\boldsymbol{r}\big)=0$.

Alternatively, we can evaluate the position operator by taking the continuum limit of the discrete-$\boldsymbol{k}$ evaluation, as was done for the cell-periodic operator, as follows:
\begin{equation}\label{diff ks}
\begin{split}
    &\langle \psi_{n'\boldsymbol{k}'}|\hat{\boldsymbol{r}}|\psi_{n\boldsymbol{k}''}\rangle
    =
    \int d\boldsymbol{r}\psi_{n'\boldsymbol{k}'}^*(\boldsymbol{r})\boldsymbol{r}\psi_{n\boldsymbol{k}''}(\boldsymbol{r})
    =
    \int d\boldsymbol{r} e^{-i\boldsymbol{k}'\cdot \boldsymbol{r}}u_{n'\boldsymbol{k}'}^*(\boldsymbol{r})
    \big(-i\boldsymbol{\partial_{k''}}e^{i\boldsymbol{k}''\cdot \boldsymbol{r}}\big)u_{n\boldsymbol{k}''}(\boldsymbol{r})\\
    &=V_\text{B.Z.}\int_\text{cell} d\boldsymbol{r}\bigg(
     \delta(\boldsymbol{k}'-\boldsymbol{k}'')e^{-i(\boldsymbol{k}'-\boldsymbol{k}'')\cdot \boldsymbol{r}}
    u_{n'\boldsymbol{k}'}^*(\boldsymbol{r})i\boldsymbol{\partial_{k''}}u_{n\boldsymbol{k}''}(\boldsymbol{r})
    -i\boldsymbol{\partial_{k''}}\Big(\delta(\boldsymbol{k}'-\boldsymbol{k}'')e^{-i(\boldsymbol{k}'-\boldsymbol{k}'')\cdot \boldsymbol{r}}
    u_{n'\boldsymbol{k}'}^*(\boldsymbol{r})u_{n\boldsymbol{k}''}(\boldsymbol{r})\Big)
    \bigg)\\
    &=V_\text{B.Z.}\int_\text{cell} d\boldsymbol{r}\bigg(
     \delta(\boldsymbol{k}'-\boldsymbol{k}'')
    e^{-i(\boldsymbol{k}'-\boldsymbol{k}'')\cdot \boldsymbol{r}}u_{n'\boldsymbol{k}'}^*(\boldsymbol{r})i\boldsymbol{\partial_{k''}}u_{n\boldsymbol{k}''}(\boldsymbol{r})
    -i\boldsymbol{\partial_{k''}}\Big(
    \delta(\boldsymbol{k}'-\boldsymbol{k}'')\psi_{n'\boldsymbol{k}'}^*(\boldsymbol{r})\psi_{n\boldsymbol{k}''}(\boldsymbol{r})
    \Big)
    \bigg)\\
    &=V_\text{B.Z.}\int_\text{cell} d\boldsymbol{r}\bigg(
     \delta(\boldsymbol{k}'-\boldsymbol{k}'')
    e^{-i(\boldsymbol{k}'-\boldsymbol{k}'')\cdot \boldsymbol{r}}u_{n'\boldsymbol{k}'}^*(\boldsymbol{r})u_{n\boldsymbol{k}''}(\boldsymbol{r})\boldsymbol{r}
    -\psi_{n'\boldsymbol{k}'}^*(\boldsymbol{r})\psi_{n\boldsymbol{k}''}(\boldsymbol{r})i\boldsymbol{\partial_{k''}}\delta(\boldsymbol{k}'-\boldsymbol{k}'')
    \bigg).
\end{split}
\end{equation}
This expression involves a differentiated Dirac delta function, which is ill-defined when $\boldsymbol{k}'=\boldsymbol{k}''$.
Moreover, even if one further evaluates this expression by regularizing the divergence---replacing $V_{\text{B.Z.}}\delta(\boldsymbol{k}'-\boldsymbol{k}'')$ with $N\delta_{\boldsymbol{k}'\boldsymbol{k}''}$---the result for the same $\boldsymbol{k}$ still depends on the choice of unit cell.

In contrast to using standard Bloch states, one can evaluate the position operator by following the proposed three rules, which provide a regularization scheme that guarantees a result independent of the choice of unit cell and satisfying the Hermitian conjugate relation.
Specifically, the evaluation of the position operator via the proposed three rules yields the Berry connection matrix in the cell-periodic Bloch basis:
\begin{equation}\label{Explicit r}
\begin{split}
    &\lim_{N\rightarrow\infty}\frac{1}{N}\langle \Psi^{N}_{n'\boldsymbol{k}}|\hat{\boldsymbol{r}}|\Psi^{N}_{n\boldsymbol{k}}\rangle
    =\lim_{N\rightarrow \infty}\frac{1}{N}\iint d\boldsymbol{k}'d\boldsymbol{k}''\Delta_{N}(\boldsymbol{k}-\boldsymbol{k}')\Delta_{N}(\boldsymbol{k}-\boldsymbol{k}'')\langle \psi_{n'\boldsymbol{k}'}|\hat{\boldsymbol{r}}|\psi_{n\boldsymbol{k}''}\rangle\\
    &=\lim_{N\rightarrow \infty}\frac{V_\text{B.Z.}}{N}\iint d\boldsymbol{k}'d\boldsymbol{k}''\Delta_{N}(\boldsymbol{k}-\boldsymbol{k}')\Delta_{N}(\boldsymbol{k}-\boldsymbol{k}'')\Bigg(
    \int_\text{cell} d\boldsymbol{r} \delta(\boldsymbol{k}'-\boldsymbol{k}'')e^{-i(\boldsymbol{k}'-\boldsymbol{k}'')\cdot \boldsymbol{r}}
    u_{n'\boldsymbol{k}'}^*(\boldsymbol{r})i\boldsymbol{\partial_{k''}}u_{n\boldsymbol{k}''}(\boldsymbol{r})\\
    &-i\boldsymbol{\partial_{k''}}\int_\text{cell} d\boldsymbol{r}
    \delta(\boldsymbol{k}'-\boldsymbol{k}'')\psi_{n'\boldsymbol{k}'}^*(\boldsymbol{r})\psi_{n\boldsymbol{k}''}(\boldsymbol{r})
    \Bigg)\\
    &=\lim_{N\rightarrow \infty}\frac{V_\text{B.Z.}}{N}\int d\boldsymbol{k}'\Big(
    \big(\Delta_{N}(\boldsymbol{k}-\boldsymbol{k}')\big)^2[\boldsymbol{A}(\boldsymbol{k}')]_{n'n}
    +\Delta_{N}(\boldsymbol{k}-\boldsymbol{k}')i\boldsymbol{\partial_{k'}}\Delta_{N}(\boldsymbol{k}-\boldsymbol{k}')\delta_{n'n}
    \Big)=[\boldsymbol{A}(\boldsymbol{k})]_{n'n},
\end{split}
\end{equation}
where the last step used $\int d\boldsymbol{k}'\Delta_{N}(\boldsymbol{k}-\boldsymbol{k}')i\boldsymbol{\partial_{k'}}\Delta_{N}(\boldsymbol{k}-\boldsymbol{k}')=\frac{1}{2}\int d\boldsymbol{k}'i\boldsymbol{\partial_{k'}}\big(\Delta_{N}(\boldsymbol{k}-\boldsymbol{k}')\big)^{2}=0$.
This result is independent of the choice of unit cell because of its cell periodicity and satisfies the Hermitian conjugate relation; i.e., $\lim_{N\rightarrow\infty}N^{-1}\langle \Psi^{N}_{n'\boldsymbol{k}}|\hat{\boldsymbol{r}}|\Psi^{N}_{n\boldsymbol{k}}\rangle^{*}
=\lim_{N\rightarrow\infty}N^{-1}\langle \Psi^{N}_{n\boldsymbol{k}}|\hat{\boldsymbol{r}}^{\dagger}|\Psi^{N}_{n'\boldsymbol{k}}\rangle$.
Moreover, it satisfies the physical expectation that the time derivative of the intraband matrix element of the position operator corresponds to the intraband velocity:
\begin{equation}
    \frac{d}{dt}[\boldsymbol{A}(\boldsymbol{k})]_{nn}=i\frac{d}{dt}\langle u_{n\boldsymbol{k}}|\boldsymbol{\partial_{k}}u_{n\boldsymbol{k}}\rangle
    =\hbar^{-1}\bigg(-\big(\langle u_{n\boldsymbol{k}}|\hat{H}(\boldsymbol{k})\big)|\boldsymbol{\partial_{k}}u_{n\boldsymbol{k}}\rangle
    +\langle u_{n\boldsymbol{k}}|\boldsymbol{\partial_{k}}\big(\hat{H}(\boldsymbol{k})|u_{n\boldsymbol{k}}\rangle\big)\bigg)
    =\hbar^{-1}\boldsymbol{\partial_{k}}E_{n}(\boldsymbol{k}).
\end{equation}

\subsection{Applying three rules to evaluate the general result}\label{problem diff Dirac}
When the position operator is absent from a product chain, no differentiated Dirac delta functions appear in the evaluation, rendering the procedure straightforward, as demonstrated in Eq.~\eqref{O eval rule 2 explain}.
Thus, we focus exclusively on evaluating operator product chains that contain a single position operator, specifically of the form $\hat{O}\hat{r}\hat{Q}$, within a one-dimensional system.

Applying the proposed rules, we evaluate the product chain $\hat{O}\hat{r}\hat{Q}$ by inserting the completeness relation for Bloch states:
\begin{equation}
\begin{split}
    &\lim_{N\rightarrow\infty}\frac{1}{N}\langle \Psi^{N}_{n'k}|\hat{O}\hat{r}\hat{Q}|\Psi^{N}_{nk}\rangle\\
    &=\lim_{N\rightarrow \infty}\frac{1}{N}\iint dk'dk''\Delta_{N}(k-k')\Delta_{N}(k-k'')\sum_{m',m}\iint\frac{dk_1dk_2}{V_\text{B.Z.}^2}
    \langle\psi_{n'k'}|\hat{O}|\psi_{m'k_1}\rangle
    \langle \psi_{m'k_1}|\hat{r}|\psi_{mk_2}\rangle
    \langle\psi_{mk_2}|\hat{Q} |\psi_{nk''}\rangle\\
    &=\big[[O(k)][A(k)][Q(k)]\big]_{n'n}
    +\sum_{m}\frac{1}{2}\Big(
    [O(k)]_{n'm}\big(i\partial_{k}[Q(k)]_{mn}\big)
    -\big(i\partial_{k}[O(k)]_{n'm}\big)[Q(k)]_{mn}
    \Big).
\end{split}
\end{equation}
Here, the first and second terms in the final expression correspond to the two terms in Eq.~\eqref{r in Bloch}, respectively.
The first term evaluates to:
\begin{equation}
\begin{split}
    &\lim_{N\rightarrow \infty}\frac{1}{N}\sum_{m',m}\iint dk'dk''\Delta_{N}(k-k')\Delta_{N}(k-k'')
    \iint\frac{ dk_1dk_2}{V_\text{B.Z.}^2}
    \langle\psi_{n'k'}|\hat{O}|\psi_{m'k_1}\rangle
    \langle\psi_{mk_2}|\hat{Q} |\psi_{nk''}\rangle\\
    &\times\int_\text{cell} dr \delta(k_1-k_2)e^{-i(k_1-k_2) r}
    u_{m'k_1}^*(r)i\partial_{k_2}u_{m k_2}(r)\\
    &=\lim_{N\rightarrow \infty}\frac{V_\text{B.Z.}}{N}\sum_{m',m}\int dk'
    \big(\Delta_{N}(k-k')\big)^2 [O(k')]_{n'm'}[A(k')]_{m'm}[Q(k')]_{mn}
    =\big[[O(k)][A(k)][Q(k)]\big]_{n'n},
\end{split}
\end{equation}
while the second term evaluates to:
\begin{equation}\label{Calcul Gen explicit for diff part}
\begin{split}
    &\lim_{N\rightarrow \infty}\frac{1}{N}\sum_{m',m}\iint dk'dk''\Delta_{N}(k-k')\Delta_{N}(k-k'')
    \iint\frac{ dk_1dk_2}{V_\text{B.Z.}^2}
    \langle\psi_{n'k'}|\hat{O}|\psi_{m'k_1}\rangle
    \langle\psi_{mk_2}|\hat{Q} |\psi_{nk''}\rangle\\
    &\quad\times\bigg(
    -i\partial_{k_2}\int_\text{cell} dr\delta(k_1-k_2)e^{-i(k_1-k_2)r}
    u_{m'k_1}^*(r)u_{mk_2}(r)
    \bigg)\\
    &=\lim_{N\rightarrow \infty}\frac{V_\text{B.Z.}}{N}\sum_{m',m}\iint dk'dk''\Delta_{N}(k-k')\Delta_{N}(k-k'') [O(k')]_{n'm'}[Q(k'')]_{mn}\\
    &\quad\times\bigg(-i\partial_{k''}\int_\text{cell} dr\delta(k'-k'')e^{-i(k'-k'')r}
    u_{m'k'}^*(r)u_{mk''}(r)
    \bigg)\\
    &=\lim_{N\rightarrow \infty}\frac{V_\text{B.Z.}}{2N}\sum_{m}\int dk'\Big(
    \Delta_{N}(k-k')[O(k')]_{n'm}i\partial_{k'}\big(\Delta_{N}(k-k')[Q(k')]_{mn}\big)\\
    &\quad-i\partial_{k'}\big(\Delta_{N}(k-k')[O(k')]_{n'm}\big)\Delta_{N}(k-k')[Q(k')]_{mn}
    \Big)\\
    &=\sum_{m}\frac{1}{2}\Big(
    [O(k)]_{n'm}\big(i\partial_{k}[Q(k)]_{mn}\big)
    -\big(i\partial_{k}[O(k)]_{n'm}\big)[Q(k)]_{mn}
    \Big).
\end{split}
\end{equation}
We distribute the derivative symmetrically acting on the factors toward the bra and ket sides, systematically eliminating the differentiated $\Delta_N$ functions.
This outcome is exactly reproduced by applying a one-sided derivative utilizing integration by parts with a vanishing surface term, as demonstrated in Eq.~\eqref{Explicit r}, or by changing the variable of the derivative acting on $\Delta_N$ functions.

For the last comment in this section, we extend the application of the proposed rules to evaluate $\lim_{N\rightarrow \infty}\langle f^{N}|\hat{O}\hat{r}\hat{Q}|g^{N}\rangle$ within the generalized states, $|f^{N}\rangle=\int \frac{dk}{V_{\text{B.Z.}}} |\Psi^{N}_{n'k}\rangle f(k)$ and $|g^{N}\rangle=\int \frac{dk}{V_{\text{B.Z.}}}|\Psi^{N}_{nk}\rangle g(k)$.
In the presence of a nonzero surface term over the Brillouin zone originating from either matrix element $[O(k)]$ or $[Q(k)]$---such as the Berry connection in a Chern insulator---the proposed rules accompany the ambiguity resulting from a remaining nonzero surface term emerging otherwise.
Specifically, the evaluation from the symmetrically distributed derivatives yields 
\begin{equation}
\begin{split}
    &\int \frac{dk}{V_{\text{B.Z.}}}\bigg(
    f^*(k) \big[[O(k)][A(k)][Q(k)]\big]_{n'n} g(k)\\
    &+\frac{1}{2}\sum_{m}\Big(
    f^*(k) [O(k)]_{n'm}i\partial_{k}\big([Q(k)]_{mn} g(k)\big)-
    i\partial_{k}\big( f^*(k) [O(k)]_{n'm}\big)[Q(k)]_{mn} g(k)
    \Big)
    \bigg),
\end{split}
\end{equation}
but that from the right-sided derivative produces an additional term $-\int \frac{dk}{2V_{\text{B.Z.}}}i\partial_{k}\Big( f^*(k) \big[[O(k)][Q(k)]\big]_{n'n} g(k)\Big)$.
However, this ambiguity problem is inherently unavoidable because it is rooted in the ambiguity of the nonzero surface term in the calculation $\iint dkdk' f(k) g(k') \partial_k\delta(k-k')$, which can result in either $\int dk f(k)\partial_kg(k)$ or $-\int dk' g(k') \partial_{k'} f(k')$.
Therefore, to ensure an unambiguous evaluation and to satisfy the Hermitian conjugate relation---$\lim_{N\rightarrow \infty}\langle f^{N}|\hat{O}\hat{r}\hat{Q}|g^{N}\rangle = \lim_{N\rightarrow \infty}\langle g^{N}|\hat{Q}\hat{r}\hat{O}|f^{N}\rangle^*$---we adopt the use of symmetrically distributed derivatives at least in the presence of a nonzero surface term over the Brillouin zone as a natural extension.

\section{Self-rotation and local circulation}\label{LC SR Appendix}
Two distinct formulations exist for evaluating the orbital contribution to the magnetism, i.e., the sum of the local and itinerant circulations~\cite{Thonhauser2005PhysRevLett.95.137205, Ceresoli2006PhysRevB.74.024408} and the self-rotation when accounting for the modification of the density of states~\cite{Xiao2005PhysRevLett.95.137204, Xiao2010RevModPhys.82.1959}.
Although their results in total are identical, their conventional evaluations of the orbital moment operator---namely, the self-rotation and local circulation---show a discrepancy.
However, these evaluations should not differ at the operator level because the local circulation evaluated in the Wannier representation is merely the self-rotation result obtained by substituting the distribution $\mathcal{W}(\boldsymbol{k})$ with $e^{i\boldsymbol{k}\cdot\boldsymbol{R}}$ and summing over the lattice vectors $\boldsymbol{R}$.
In what follows, we demonstrate that these two approaches yield identical results under rigorous evaluation, although the methodologies employed to remove gauge dependence for the self-rotation and local circulation are distinctly different.

\subsection{Missing terms in the self-rotation}\label{Missing term in SR}
In the literature incorporating self-rotation~\cite{Xiao2005PhysRevLett.95.137204, Xiao2010RevModPhys.82.1959}, the associated formulations originate from the expression derived in Ref.~\cite{Chang1996PhysRevB.53.7010}:
\begin{equation}\label{SR whole term}
\begin{split}
    &\langle \mathcal{W}|(\hat{\boldsymbol{r}}-\boldsymbol{r}_{\mathcal{W}})\times \hat{\boldsymbol{v}}|\mathcal{W}\rangle
    =\int\frac{d\boldsymbol{k}}{V_{\text{B.Z.}}}|\tilde{\mathcal{W}}(\boldsymbol{k})|^2\frac{i}{\hbar}
    \langle \boldsymbol{\partial_{k}}u_{n\boldsymbol{k}}|\times (\hat{H}(\boldsymbol{k})-E_{n}(\boldsymbol{k}))|\boldsymbol{\partial_{k}} u_{n\boldsymbol{k}}\rangle\\
    &+\int\frac{d\boldsymbol{k}}{V_{\text{B.Z.}}}\bigg(
    |\tilde{\mathcal{W}}(\boldsymbol{k})|^2
    \langle i\boldsymbol{\partial_{k}}u_{n\boldsymbol{k}}|u_{n\boldsymbol{k}}\rangle
    +\big(-i\boldsymbol{\partial_{k}}\tilde{\mathcal{W}}^{*}(\boldsymbol{k})\big)\tilde{\mathcal{W}}(\boldsymbol{k})
    \bigg)\times \boldsymbol{v}_{n}(\boldsymbol{k})\\
    &= \int\frac{d\boldsymbol{k}}{V_{\text{B.Z.}}}|\mathcal{W}(\boldsymbol{k})|^2\frac{i}{\hbar}
    \langle \boldsymbol{\partial_{k}}u_{n\boldsymbol{k}}|\times (\hat{H}(\boldsymbol{k})-E_{n}(\boldsymbol{k}))|\boldsymbol{\partial_{k}} u_{n\boldsymbol{k}}\rangle\\
    &+\int\frac{d\boldsymbol{k}}{V_{\text{B.Z.}}}\bigg(
    |\mathcal{W}(\boldsymbol{k})|^2\langle i\boldsymbol{\partial_{k}}u_{n\boldsymbol{k}}|u_{n\boldsymbol{k}}\rangle
    +\big(-i\boldsymbol{\partial_{k}}\mathcal{W}^{*}(\boldsymbol{k})\big)\mathcal{W}(\boldsymbol{k})
    -|\mathcal{W}(\boldsymbol{k})|^2\boldsymbol{r}_{\mathcal{W}}
    \bigg)
    \times \boldsymbol{v}_{n}(\boldsymbol{k}),
\end{split}
\end{equation}
where $\tilde{\mathcal{W}}(\boldsymbol{k})=\mathcal{W}(\boldsymbol{k})e^{i\boldsymbol{k}\cdot\boldsymbol{r}_{\mathcal{W}}}$, $\int\frac{d\boldsymbol{k}}{V_{\text{B.Z.}}}|\mathcal{W}(\boldsymbol{k})|^2=1$, $|\mathcal{W}\rangle=\int\frac{d\boldsymbol{k}}{V_{\text{B.Z.}}}\mathcal{W}(\boldsymbol{k})|\psi_{n\boldsymbol{k}}\rangle$, and the position expectation value is given by
\begin{equation}
\begin{split}
    &\boldsymbol{r}_{\mathcal{W}}
    =\langle \mathcal{W}|\hat{\boldsymbol{r}}|\mathcal{W}\rangle
    =\int\frac{d\boldsymbol{k}}{V_{\text{B.Z.}}}\bigg(
    |\mathcal{W}(\boldsymbol{k})|^2\langle i\boldsymbol{\partial_{k}}u_{n\boldsymbol{k}}|u_{n\boldsymbol{k}}\rangle
    +\big(-i\boldsymbol{\partial_{k}}\mathcal{W}^{*}(\boldsymbol{k})\big)\mathcal{W}(\boldsymbol{k})
    \bigg).
\end{split}
\end{equation}
To recover the conventional form of the self-rotation, as expressed in
\begin{equation}\label{SR in literatures}
    \langle \mathcal{W}|(\hat{\boldsymbol{r}}-\boldsymbol{r}_{\mathcal{W}})\times \hat{\boldsymbol{v}}|\mathcal{W}\rangle
    =\int\frac{d\boldsymbol{k}}{V_{\text{B.Z.}}}
    |\mathcal{W}(\boldsymbol{k})|^2
    \frac{i}{\hbar}\langle \boldsymbol{\partial_{k}}u_{n\boldsymbol{k}}|\times (\hat{H}(\boldsymbol{k})-E_{n}(\boldsymbol{k}))|\boldsymbol{\partial_{k}} u_{n\boldsymbol{k}}\rangle,
\end{equation}
the assumption introduced in Eq.~(B6) of Ref.~\cite{Chang1996PhysRevB.53.7010} must be invoked:
\begin{equation}\label{SR error}
\begin{split}
    \int\frac{d\boldsymbol{k}}{V_{\text{B.Z.}}}\bigg(
    |\mathcal{W}(\boldsymbol{k})|^2\langle i\boldsymbol{\partial_{k}}u_{n\boldsymbol{k}}|u_{n\boldsymbol{k}}\rangle
    +\big(-i\boldsymbol{\partial_{k}}\mathcal{W}^{*}(\boldsymbol{k})\big)\mathcal{W}(\boldsymbol{k})
    -|\mathcal{W}(\boldsymbol{k})|^2\boldsymbol{r}_{\mathcal{W}}
    \bigg)
    \times \boldsymbol{v}_{n}(\boldsymbol{k})
    =0.
\end{split}
\end{equation}
However, the validity of Eq.~\eqref{SR error} requires careful reexamination under rigorous evaluation.
In general, the integral $\int d\boldsymbol{k}\big(f(\boldsymbol{k})-\bar{f}\big)g(\boldsymbol{k})p(\boldsymbol{k})\neq 0$ for $\bar{f}= \int d\boldsymbol{k}f(\boldsymbol{k})p(\boldsymbol{k})$; mapping this general mathematical relation back to Eq.~\eqref{SR error} involves the substitutions $f(\boldsymbol{k})=\langle i\boldsymbol{\partial_{k}}u_{n\boldsymbol{k}}|u_{n\boldsymbol{k}}\rangle-\big(\mathcal{W}^{*}(\boldsymbol{k})\big)^{-1}i\boldsymbol{\partial_{k}}\mathcal{W}^{*}(\boldsymbol{k})$, $\bar{f}=\boldsymbol{r}_{\mathcal{W}}$, $g(\boldsymbol{k})=\boldsymbol{v}_{n}(\boldsymbol{k})$, and $p(\boldsymbol{k})=|\mathcal{W}(\boldsymbol{k})|^2/V_{\text{B.Z.}}$.

By evaluating the complete expression in Eq.~\eqref{SR whole term} without invoking the assumption from Eq.~\eqref{SR error}, we establish a formal connection between the self-rotation and the local circulation.
This connection is achieved by applying a transformation from the wave packet $|\mathcal{W}\rangle$ to the Wannier function $|W_{n\boldsymbol{R}}\rangle$ via the limit $\mathcal{W}(\boldsymbol{k})\rightarrow e^{i\boldsymbol{k}\cdot \boldsymbol{R}}$, which directly maps $|\mathcal{W}\rangle$ onto $|W_{n\boldsymbol{R}}\rangle$.
Consequently, the evaluation yields the self-rotation as the Brillouin zone integration of the orbital moment, expressed as:
\begin{equation}\label{SR correct}
\begin{split}
    &\langle W_{n\boldsymbol{R}}|\big(\hat{\boldsymbol{r}}-\boldsymbol{r}_{c,n}(\boldsymbol{R})\big)\times \hat{\boldsymbol{v}}|W_{n\boldsymbol{R}} \rangle
    =\int\frac{d\boldsymbol{k}}{V_{\text{B.Z.}}}\bigg(
    \frac{i}{\hbar}\langle \boldsymbol{\partial_{k}}u_{n\boldsymbol{k}}|\times (\hat{H}(\boldsymbol{k})-E_{n}(\boldsymbol{k}))|\boldsymbol{\partial_{k}} u_{n\boldsymbol{k}}\rangle
    +\Big([\boldsymbol{A}(\boldsymbol{k})]_{nn}+\boldsymbol{R}\Big)\times \boldsymbol{v}_{n}(\boldsymbol{k})\bigg)\\
    &=\int\frac{d\boldsymbol{k}}{V_{\text{B.Z.}}}\bigg(
    \frac{i}{\hbar}\langle \boldsymbol{\partial_{k}}u_{n\boldsymbol{k}}|\times (\hat{H}(\boldsymbol{k})-E_{n}(\boldsymbol{k}))|\boldsymbol{\partial_{k}} u_{n\boldsymbol{k}}\rangle
    +[\boldsymbol{A}(\boldsymbol{k})]_{nn}\times \boldsymbol{v}_{n}(\boldsymbol{k})\bigg)
    =\int\frac{d\boldsymbol{k}}{V_{\text{B.Z.}}}
    \lim_{N\rightarrow \infty}\frac{1}{N}\langle \Psi^{N}_{n\boldsymbol{k}}|2\hat{\boldsymbol{m}}|\Psi^{N}_{n\boldsymbol{k}}\rangle\\
    &=\int\frac{d\boldsymbol{k}}{V_{\text{B.Z.}}}2[\boldsymbol{m}(\boldsymbol{k})]_{nn},
\end{split}
\end{equation}
where $\boldsymbol{r}_{c,n}(\boldsymbol{R})=\langle W_{n\boldsymbol{R}}|\hat{\boldsymbol{r}}|W_{n\boldsymbol{R}} \rangle$, and the factor 2 arises from the different conventions of the orbital moment in Eq.~\eqref{Orbital moment equal band Eval}.
Since the outcome remains independent of $\boldsymbol{R}$, applying a normalized summation over the lattice vectors---as performed in the evaluation of the local circulation---does not alter the result.

\subsection{Missing terms in the local circulation and the Wannier function regularization}\label{Missing term in LC}
In this subsection, we analyze the issues associated with using Wannier functions to evaluate the position operator.
We demonstrate that the $\Delta_N$-regularized Wannier states resolve these issues, analogous to the use of $\Delta_N$-regularized Bloch states.
By applying this regularization, we prove that the evaluation of the local circulation becomes identical to that of the orbital moment, and show that the discrepancy between our evaluation and previously known results is physically significant.

\subsubsection{Issues in the Wannier representation}\label{Wannier function comments}
Before examining the local circulation in the Wannier representation, we must address an inherent problem within the Wannier representation itself.
Although the position operator appears well-defined in this representation (see Eq.~\eqref{r eval in Wannier})---unlike in the Bloch representation---its direct application leads to inconsistencies.
The issue does not originate from the evaluation of the position operator matrix elements themselves: $\langle W_{n'\boldsymbol{R}'}|\hat{\boldsymbol{r}}|W_{n\boldsymbol{R}}\rangle=\int\frac{d\boldsymbol{k}}{V_{\text{B.Z.}}}e^{i\boldsymbol{k}\cdot(\boldsymbol{R}'-\boldsymbol{R})}[\boldsymbol{A}(\boldsymbol{k})]_{n'n}+\delta_{n'n}\int\frac{d\boldsymbol{k}}{V_{\text{B.Z.}}}e^{i\boldsymbol{k}\cdot\boldsymbol{R}'}i\boldsymbol{\partial_{k}} e^{-i\boldsymbol{k}\cdot\boldsymbol{R}}=[\boldsymbol{A}(\boldsymbol{R}'-\boldsymbol{R})]_{n'n}+\delta_{n'n}\delta_{\boldsymbol{R}'\boldsymbol{R}}\boldsymbol{R}$, where the Wannier function is defined as $W_{n\boldsymbol{R}}(\boldsymbol{r})=\langle \boldsymbol{r}|W_{n\boldsymbol{R}}\rangle=\int \frac{d\boldsymbol{k}}{V_{\text{B.Z.}}}e^{-i\boldsymbol{k}\cdot\boldsymbol{R}}\psi_{n\boldsymbol{k}}(\boldsymbol{r})$, $n$ is the band index, and $\boldsymbol{R}$ denotes the lattice vectors~\cite{Vanderbilt2018Book}.
However, treating the lattice vector $\boldsymbol{R}$ as an isolated multiplier outside of the exponential phase factor introduces two critical problems: it breaks the Born--von Karman boundary conditions imposed on the set of Wannier functions, and it effectively neglects surface terms due to an implicit integration by parts.

The first problem manifests in the unbounded summation over the lattice vectors, $\sum_{\boldsymbol{R}}\boldsymbol{R}$.
Specifically, the macroscopic translation invariance dictated by the Born--von Karman boundary conditions---which requires $\lim_{N\rightarrow\infty}N^{-1}\sum_{\boldsymbol{R}}\langle W_{n'\boldsymbol{R}}|\hat{\boldsymbol{r}}|W_{n\boldsymbol{R}}\rangle=\lim_{N\rightarrow\infty}N^{-1}\sum_{\boldsymbol{R}}\langle W_{n'(\boldsymbol{R}+\tilde{\boldsymbol{R}})}|\hat{\boldsymbol{r}}|W_{n(\boldsymbol{R}+\tilde{\boldsymbol{R}})}\rangle$---is violated as follows: 
\begin{equation}
\begin{split}
    &\lim_{N\rightarrow\infty}\frac{1}{N}\sum_{\boldsymbol{R}}\langle W_{n\boldsymbol{R}}|\hat{\boldsymbol{r}}|W_{n\boldsymbol{R}}\rangle
    =[\boldsymbol{A}(\boldsymbol{R}=0)]_{nn}+\lim_{N\rightarrow\infty}\frac{1}{N}\sum_{\boldsymbol{R}} \boldsymbol{R},
    \\
    &\lim_{N\rightarrow\infty}\frac{1}{N}\sum_{\boldsymbol{R}}\langle W_{n(\boldsymbol{R}+\tilde{\boldsymbol{R}})}|\hat{\boldsymbol{r}}|W_{n(\boldsymbol{R}+\tilde{\boldsymbol{R}})}\rangle
    =[\boldsymbol{A}(\boldsymbol{R}=0)]_{nn}+\lim_{N\rightarrow\infty}\frac{1}{N}\sum_{\boldsymbol{R}} (\boldsymbol{R}+\tilde{\boldsymbol{R}})\\
    &=\lim_{N\rightarrow\infty}\frac{1}{N}\sum_{\boldsymbol{R}}\langle W_{n\boldsymbol{R}}|\hat{\boldsymbol{r}}|W_{n\boldsymbol{R}}\rangle+\tilde{\boldsymbol{R}}
    \neq \lim_{N\rightarrow\infty}\frac{1}{N}\sum_{\boldsymbol{R}}\langle W_{n\boldsymbol{R}}|\hat{\boldsymbol{r}}|W_{n\boldsymbol{R}}\rangle.
\end{split}
\end{equation}
This discrepancy contradicts the fundamental periodic nature of the crystal.
Just as the macroscopic sum $N^{-1}\sum_{n=1}^{N}e^{2\pi in/N}$ must identically equal $N^{-1}\sum_{n=1}^{N}e^{2\pi i(n+1)/N}$ over a closed ring, the sum over Wannier centers should be invariant under a global lattice shift.
However, the emergence of the bare $\boldsymbol{R}$ term explicitly breaks this translational invariance, violating the Born--von Karman boundary conditions.
Consequently, the problematic summation over lattice vectors in the Wannier representation is deeply connected to the ill-defined nature of the position operator in the Bloch representation.

The second problem becomes explicit when evaluating operators that are not surface-term-free over the Brillouin zone, such as the commutator $\langle W_{n\boldsymbol{R}}|\{\hat{r}^{i},\hat{r}^{j}\}_{-}|W_{n\boldsymbol{R}}\rangle$ in a topologically nontrivial insulator.
A similar contradiction arises for the cross product $\langle W_{n\boldsymbol{R}'}|\hat{\boldsymbol{r}}\times \hat{\boldsymbol{r}}|W_{n\boldsymbol{R}}\rangle\big|_{\boldsymbol{R}'=\boldsymbol{R}}$.
Since $\hat{\boldsymbol{r}}\times\hat{\boldsymbol{r}}=0$ identically, this expectation value must vanish.
However, a naive computation in the Wannier representation for a Chern insulator yields a nonzero result:
\begin{equation}\label{Commutation rel broken 2}
\begin{split}
    &\langle W_{n\boldsymbol{R}'}|\hat{\boldsymbol{r}}\times \hat{\boldsymbol{r}}|W_{n\boldsymbol{R}}\rangle\Big|_{\boldsymbol{R}'=\boldsymbol{R}}
    =\sum_{m,\tilde{\boldsymbol{R}}}\Big(
    \langle W_{n\boldsymbol{R}'}|\hat{\boldsymbol{r}}|W_{m\tilde{\boldsymbol{R}}}\rangle \times \langle W_{m\tilde{\boldsymbol{R}}}|\hat{\boldsymbol{r}}|W_{n\boldsymbol{R}}\rangle
    \Big)\bigg|_{\boldsymbol{R}'=\boldsymbol{R}}\\
    &=\int\frac{d\boldsymbol{k}}{V_{\text{B.Z.}}}e^{i\boldsymbol{k}\cdot(\boldsymbol{R}'-\boldsymbol{R})}\Big(
    \big[[\boldsymbol{A}(\boldsymbol{k})]\times [\boldsymbol{A}(\boldsymbol{k})]\big]_{nn}+(\boldsymbol{R}'-\boldsymbol{R})\times [\boldsymbol{A}(\boldsymbol{k})]_{nn}
    +\boldsymbol{R}'\times \boldsymbol{R}
    \Big)\bigg|_{\boldsymbol{R}'=\boldsymbol{R}}\\
    &=\int\frac{d\boldsymbol{k}}{V_{\text{B.Z.}}}
    \big[[\boldsymbol{A}(\boldsymbol{k})]\times [\boldsymbol{A}(\boldsymbol{k})]\big]_{nn}
    \neq 0,
\end{split}
\end{equation}
yielding an identical result for the evaluation:
\begin{equation}
\begin{split}
    &\lim_{N\rightarrow \infty}\frac{1}{N}\sum_{\tilde{\boldsymbol{R}}}\langle W_{n(\boldsymbol{R}'+\tilde{\boldsymbol{R}})}|\hat{\boldsymbol{r}}\times \hat{\boldsymbol{r}}|W_{n(\boldsymbol{R}+\tilde{\boldsymbol{R}})}\rangle\big|_{\boldsymbol{R}'=\boldsymbol{R}}
    =\int\frac{d\boldsymbol{k}}{V_{\text{B.Z.}}}
    \big[[\boldsymbol{A}(\boldsymbol{k})]\times [\boldsymbol{A}(\boldsymbol{k})]\big]_{nn}
    \neq 0.
\end{split}
\end{equation}

\subsubsection{Regularization of the Wannier function}\label{Regularization of Wannier functions}
To resolve the aforementioned issues, we propose substituting the conventional Wannier states with $\Delta_N$-regularized Wannier states $|\mathfrak{W}^{N}_{n\boldsymbol{R}}\rangle$ in Sec.~\ref{BSection 4}.
Employing this substitution establishes the evaluation of the operator in the Wannier representation, $\lim_{N\rightarrow\infty}\langle \mathfrak{W}^{N}_{n'\boldsymbol{R}'}|\hat{O}|\mathfrak{W}^{N}_{n\boldsymbol{R}}\rangle$, expressed as follows:
\begin{equation}
\begin{split}
    &\lim_{N\rightarrow\infty}
    \langle \mathfrak{W}^{N}_{n'\boldsymbol{R}'}|\hat{O}|\mathfrak{W}^{N}_{n\boldsymbol{R}}\rangle=
    \lim_{N\rightarrow\infty}\iint\frac{d\tilde{\boldsymbol{k}}d\tilde{\boldsymbol{k}}'}{V_{\text{B.Z.}}^2}e^{i\tilde{\boldsymbol{k}}'\cdot\boldsymbol{R}'-i\tilde{\boldsymbol{k}}\cdot\boldsymbol{R}}
    \langle \Psi^{N}_{n'\tilde{\boldsymbol{k}}'}|\hat{O}|\Psi^{N}_{n\tilde{\boldsymbol{k}}}\rangle.
\end{split}
\end{equation}
Furthermore, this regularization framework establishes four key relations, with the detailed derivations provided in Appendix~\ref{Wannier rep regularization}:
\begin{enumerate}
    \item $\lim_{N\rightarrow\infty}\langle \mathfrak{W}^{N}_{n'\boldsymbol{R}'}|\hat{O}|\mathfrak{W}^{N}_{n\boldsymbol{R}}\rangle=\langle W_{n'\boldsymbol{R}'}|\hat{O}|W_{n\boldsymbol{R}}\rangle$.
    \item $\lim_{N\rightarrow\infty}N^{-1}\sum_{\boldsymbol{R}}\langle \mathfrak{W}^{N}_{n'\boldsymbol{R}}|\hat{O}|\mathfrak{W}^{N}_{n\boldsymbol{R}}\rangle=\lim_{N\rightarrow\infty}N^{-1}\sum_{\boldsymbol{R}}\langle W_{n'\boldsymbol{R}}|\hat{O}|W_{n\boldsymbol{R}}\rangle$.
    \item $\lim_{N\rightarrow\infty}\langle \mathfrak{W}^{N}_{n'\boldsymbol{R}'}|\hat{r}|\mathfrak{W}^{N}_{n\boldsymbol{R}}\rangle=\langle W_{n'\boldsymbol{R}'}|\hat{r}|W_{n\boldsymbol{R}}\rangle$.
    \item $\lim_{N\rightarrow\infty}N^{-1}\sum_{\boldsymbol{R}}\langle \mathfrak{W}^{N}_{n'\boldsymbol{R}}|\hat{r}|\mathfrak{W}^{N}_{n\boldsymbol{R}}\rangle \neq \lim_{N\rightarrow\infty}N^{-1}\sum_{\boldsymbol{R}}\langle W_{n'\boldsymbol{R}}|\hat{r}|W_{n\boldsymbol{R}}\rangle$, but $\lim_{N\rightarrow\infty}N^{-1}\sum_{\boldsymbol{R}}\langle \mathfrak{W}^{N}_{n'\boldsymbol{R}}|\hat{r}|\mathfrak{W}^{N}_{n\boldsymbol{R}}\rangle= \lim_{N\rightarrow\infty}N^{-1}\sum_{\boldsymbol{R}}\langle \mathfrak{W}^{N}_{n'(\boldsymbol{R}+\tilde{\boldsymbol{R}})}|\hat{r}|\mathfrak{W}^{N}_{n(\boldsymbol{R}+\tilde{\boldsymbol{R}})}\rangle$.
\end{enumerate}
Here, we focus on demonstrating that $\lim_{N\rightarrow \infty}N^{-1}\sum_{\tilde{\boldsymbol{R}}}\langle \mathfrak{W}^{N}_{n'(\boldsymbol{R}'+\tilde{\boldsymbol{R}})}|\hat{\boldsymbol{r}}\times \hat{\boldsymbol{r}}|\mathfrak{W}^{N}_{n(\boldsymbol{R}+\tilde{\boldsymbol{R}})}\rangle=0$ by leveraging the $\Delta_N$-regularization alongside the established relations.
This result is derived by proving that the commutator $\{\hat{r}^{i},\hat{r}^{j}\}_{-}$ evaluates to zero, as explicitly detailed below:
\begin{equation}\label{R'R Wannier r times r calcul}
\begin{split}
    &\lim_{N\rightarrow\infty}\frac{1}{N}\sum_{\tilde{\boldsymbol{R}}}
    \langle \mathfrak{W}^{N}_{n'(\boldsymbol{R}'+\tilde{\boldsymbol{R}})}|\{\hat{r}^{i},\hat{r}^{j}\}_{-}|\mathfrak{W}^{N}_{n(\boldsymbol{R}+\tilde{\boldsymbol{R}})}\rangle
    =\lim_{N\rightarrow\infty}\frac{1}{N}\int\frac{d\boldsymbol{k}}{V_{\text{B.Z.}}}e^{i\boldsymbol{k}\cdot(\boldsymbol{R}'-\boldsymbol{R})}
    \langle \Psi^{N}_{n'\boldsymbol{k}}|\{\hat{r}^{i},\hat{r}^{j}\}_{-}|\Psi^{N}_{n\boldsymbol{k}}\rangle\\
    &=\lim_{N\rightarrow\infty}\frac{1}{N}\sum_{m}\int d\boldsymbol{k}e^{i\boldsymbol{k}\cdot(\boldsymbol{R}'-\boldsymbol{R})}
    \iiint \frac{d\tilde{\boldsymbol{k}} d\tilde{\boldsymbol{k}}'d\tilde{\boldsymbol{k}}''}{V_{\text{B.Z.}}}\Delta_{N}(\boldsymbol{k}-\tilde{\boldsymbol{k}}')\\
    &\quad\times\bigg(
    \Big([A^{i}(\tilde{\boldsymbol{k}}')]_{n'm}\delta(\tilde{\boldsymbol{k}}'-\tilde{\boldsymbol{k}}'')+\delta_{n'm}i\partial^{i}_{\tilde{k}'}\delta(\tilde{\boldsymbol{k}}'-\tilde{\boldsymbol{k}}'')\Big)
    \Big([A^{j}(\tilde{\boldsymbol{k}}'')]_{mn}\delta(\tilde{\boldsymbol{k}}''-\tilde{\boldsymbol{k}})+\delta_{mn}i\partial^{j}_{\tilde{k}''}\delta(\tilde{\boldsymbol{k}}''-\tilde{\boldsymbol{k}})\Big)\\
    &\quad-\Big([A^{j}(\tilde{\boldsymbol{k}}')]_{n'm}\delta(\tilde{\boldsymbol{k}}'-\tilde{\boldsymbol{k}}'')+\delta_{n'm}i\partial^{j}_{\tilde{k}'}\delta(\tilde{\boldsymbol{k}}'-\tilde{\boldsymbol{k}}'')\Big)
    \Big([A^{i}(\tilde{\boldsymbol{k}})]_{mn}\delta(\tilde{\boldsymbol{k}}''-\tilde{\boldsymbol{k}})-\delta_{mn}i\partial^{i}_{\tilde{k}}\delta(\tilde{\boldsymbol{k}}''-\tilde{\boldsymbol{k}})\Big)
    \bigg)\Delta_{N}(\boldsymbol{k}-\tilde{\boldsymbol{k}})\\
    &=\lim_{N\rightarrow\infty}\frac{1}{N}\sum_{m}\int d\boldsymbol{k}e^{i\boldsymbol{k}\cdot(\boldsymbol{R}'-\boldsymbol{R})}
    \int \frac{d\tilde{\boldsymbol{k}}}{V_{\text{B.Z.}}}\Delta_{N}(\boldsymbol{k}-\tilde{\boldsymbol{k}})\Delta_{N}(\boldsymbol{k}-\tilde{\boldsymbol{k}})\\
    &\quad\times\bigg([A^{i}(\tilde{\boldsymbol{k}})]_{n'm}[A^{j}(\tilde{\boldsymbol{k}})]_{mn}
    +\delta_{n'm}i\partial^{i}_{\tilde{k}}[A^{j}(\tilde{\boldsymbol{k}})]_{mn}
    -[A^{j}(\tilde{\boldsymbol{k}})]_{n'm}[A^{i}(\tilde{\boldsymbol{k}})]_{mn}
    -\delta_{n'm}i\partial^{j}_{\tilde{k}}[A^{i}(\tilde{\boldsymbol{k}})]_{mn}
    \bigg)=0.
\end{split}
\end{equation}

\subsubsection{Missing terms in the local circulation}
As previously pointed out, using the naive substitution $\langle W_{n'\boldsymbol{R}'}|\hat{\boldsymbol{r}}|W_{n\boldsymbol{R}}\rangle=[\boldsymbol{A}(\boldsymbol{R}'-\boldsymbol{R})]_{n'n}+\delta_{n'n}\delta_{\boldsymbol{R}'\boldsymbol{R}}\boldsymbol{R}$ implicitly omits the surface term.
Consequently, calculating the local circulation in the Wannier representation via this naive approach leads to the following incorrect result:
\begin{equation}\label{LC wrong 2}
\begin{split}
    &\langle W_{n\boldsymbol{R}}|\big(\hat{\boldsymbol{r}}-\boldsymbol{r}_{c,n}(\boldsymbol{R})\big)\times\hat{\boldsymbol{v}}|W_{n\boldsymbol{R}}\rangle=
    \frac{1}{i\hbar}\langle W_{n\boldsymbol{R}}|\hat{\boldsymbol{r}}\times\hat{H}\hat{\boldsymbol{r}}|W_{n\boldsymbol{R}}\rangle\\
    &=\frac{1}{i\hbar}\int \frac{d\boldsymbol{k}}{V_{\text{B.Z.}}}\bigg(
    [\boldsymbol{A}(\boldsymbol{k})]_{nm}\times[\boldsymbol{A}(\boldsymbol{k})]_{mn} E_{m}(\boldsymbol{k})+(\boldsymbol{R}\times\boldsymbol{R})E_{n}(\boldsymbol{k})
    +\Big(\boldsymbol{R}\times[\boldsymbol{A}(\boldsymbol{k})]_{nn}+[\boldsymbol{A}(\boldsymbol{k})]_{nn}\times\boldsymbol{R}\Big)E_{n}(\boldsymbol{k})
    \bigg)\\
    &=\frac{1}{i\hbar}\int \frac{d\boldsymbol{k}}{V_{\text{B.Z.}}}[\boldsymbol{A}(\boldsymbol{k})]_{nm}\times[\boldsymbol{A}(\boldsymbol{k})]_{mn} E_{m}(\boldsymbol{k}).
\end{split}
\end{equation}
The difference between this outcome and Eq.~\eqref{SR correct} stems from applying naive substitution to the evaluation of the velocity operator.

However, when the $\Delta_N$-regularized Wannier states introduced in Sec.~\ref{BSection 4} are incorporated, the proper local circulation calculation naturally captures the surface term:
\begin{equation}\label{LC correct}
\begin{split}
    &\lim_{N\rightarrow \infty}\frac{1}{N}\sum_{\boldsymbol{R}}\langle \mathfrak{W}^{N}_{n\boldsymbol{R}}|\hat{\boldsymbol{r}}\times\hat{\boldsymbol{v}}|\mathfrak{W}^{N}_{n\boldsymbol{R}}\rangle
    =\lim_{N\rightarrow \infty}\frac{1}{N}\sum_{\boldsymbol{R}}
    \iint \frac{d\boldsymbol{k}'d\boldsymbol{k}}{V_{\text{B.Z.}}^2}e^{i(\boldsymbol{k}'-\boldsymbol{k})\cdot\boldsymbol{R}}
    \langle \Psi^{N}_{n\boldsymbol{k}'}|\hat{\boldsymbol{r}}\times\hat{\boldsymbol{v}}|\Psi^{N}_{n\boldsymbol{k}}\rangle\\
    &=\lim_{N\rightarrow \infty}
    \int \frac{d\boldsymbol{k}}{V_{\text{B.Z.}}}
    \frac{1}{N}\langle \Psi^{N}_{n\boldsymbol{k}}|\hat{\boldsymbol{r}}\times\hat{\boldsymbol{v}}|\Psi^{N}_{n\boldsymbol{k}}\rangle
    =\int \frac{d\boldsymbol{k}}{V_{\text{B.Z.}}}2[\boldsymbol{m}(\boldsymbol{k})]_{nn}.
\end{split}
\end{equation}
This result is identical to the self-rotation result in Eq.~\eqref{SR correct}.

Finally, we discuss the difference between the conventional local circulation result in Eq.~\eqref{LC wrong 2} and that derived from $\Delta_N$-regularized Wannier states in Eq.~\eqref{LC correct}.
This difference is a gauge-dependent quantity in general, although it vanishes for a trivial insulator:
\begin{equation}
\begin{split}
    &\int\frac{d\boldsymbol{k}}{V_{\text{B.Z.}}}\bigg(
    \frac{i}{\hbar}\langle \boldsymbol{\partial_{k}}u_{n\boldsymbol{k}}|\times (\hat{H}(\boldsymbol{k})-E_{n}(\boldsymbol{k}))|\boldsymbol{\partial_{k}} u_{n\boldsymbol{k}}\rangle
    +[\boldsymbol{A}(\boldsymbol{k})]_{nn}\times \boldsymbol{v}_{n}(\boldsymbol{k})\bigg)\\
    &=\int\frac{d\boldsymbol{k}}{V_{\text{B.Z.}}}\bigg(
    \frac{i}{\hbar}\langle \boldsymbol{\partial_{k}}u_{n\boldsymbol{k}}|\times \hat{H}(\boldsymbol{k})|\boldsymbol{\partial_{k}} u_{n\boldsymbol{k}}\rangle
    -\boldsymbol{\partial_{k}}\times\Big([\boldsymbol{A}(\boldsymbol{k})]_{nn} E_{n}(\boldsymbol{k})\Big)
    \bigg)
    =\int\frac{d\boldsymbol{k}}{V_{\text{B.Z.}}}
    \frac{i}{\hbar}\langle \boldsymbol{\partial_{k}}u_{n\boldsymbol{k}}|\times \hat{H}(\boldsymbol{k})|\boldsymbol{\partial_{k}} u_{n\boldsymbol{k}}\rangle,
\end{split}
\end{equation}
where the last term (i.e., the surface term) identically vanishes, meaning that Eq.~\eqref{LC correct} is identical to the local circulation result for a trivial insulator.
However, this consequence cannot be generalized to a Chern insulator~\cite{Ceresoli2006PhysRevB.74.024408} because the last surface term depends on the gauge choice.
For example, we consider two Berry connections, $[\boldsymbol{A}^{(1)}(\boldsymbol{k})]_{nn}$ and $[\boldsymbol{A}^{(2)}(\boldsymbol{k})]_{nn}$, to cover the whole first Brillouin zone.
We introduce an infinitesimal patch $\mathcal{D}$ around the singularity of a single Berry connection $[\boldsymbol{A}^{(1)}(\boldsymbol{k})]_{nn}$.
We then apply $[\boldsymbol{A}^{(1)}(\boldsymbol{k})]_{nn}$ for the region $\text{BZ}-\mathcal{D}$ and $[\boldsymbol{A}^{(2)}(\boldsymbol{k})]_{nn}$ for the region $\mathcal{D}$.
The surface term is
\begin{equation}\label{LC filtration}
\begin{split}
    &\int\frac{d\boldsymbol{k}}{V_{\text{B.Z.}}}\boldsymbol{\partial_{k}}\times\Big(\hbar^{-1}E_{n}(\boldsymbol{k})[\boldsymbol{A}(\boldsymbol{k})]_{nn}\Big)\\
    &=\int_{\text{BZ}-\mathcal{D}}\frac{d\boldsymbol{k}}{V_{\text{B.Z.}}}
    \boldsymbol{\partial_{k}}\times\Big(\hbar^{-1}E_{n}(\boldsymbol{k})[\boldsymbol{A}^{(1)}(\boldsymbol{k})]_{nn}\Big)
    +\int_{\mathcal{D}}\frac{d\boldsymbol{k}}{V_{\text{B.Z.}}}
    \boldsymbol{\partial_{k}}\times\Big(\hbar^{-1}E_{n}(\boldsymbol{k})[\boldsymbol{A}^{(2)}(\boldsymbol{k})]_{nn}\Big)\\
    &=\int_{\partial\mathcal{D}}\frac{d\boldsymbol{k}}{V_{\text{B.Z.}}}
    \Big(\hbar^{-1}E_{n}(\boldsymbol{k})[\boldsymbol{A}^{(1)}(\boldsymbol{k})]_{nn}\Big)
    +\int_{\mathcal{D}}\frac{d\boldsymbol{k}}{V_{\text{B.Z.}}}
    \boldsymbol{\partial_{k}}\times\Big(\hbar^{-1}E_{n}(\boldsymbol{k})[\boldsymbol{A}^{(2)}(\boldsymbol{k})]_{nn}\Big)
    =E_{n}(\boldsymbol{k}_0)\mathcal{C}_{n}.
\end{split}
\end{equation}
As the patch area vanishes, the first term becomes $E_{n}(\boldsymbol{k}_0)\mathcal{C}_{n}$, where $\boldsymbol{k}_0$ is the point in the infinitesimal patch $\mathcal{D}$, and $\mathcal{C}_{n}$ is the Chern number of the $n$-th band, and the second term becomes zero as the area decreases.
Although $E_{n}(\boldsymbol{k}_0)\mathcal{C}_{n}$ might appear gauge-invariant, the location of the singular point $\boldsymbol{k}_0$ is inherently gauge-dependent~\cite{Vanderbilt2018Book}.
Therefore, we cannot extend the local circulation result in the trivial insulator to a Chern insulator~\cite{Ceresoli2006PhysRevB.74.024408}, and the correct local circulation result must contain the surface term in addition to the local circulation result $\int\frac{d\boldsymbol{k}}{V_{\text{B.Z.}}}\frac{i}{\hbar}\langle \boldsymbol{\partial_{k}}u_{n\boldsymbol{k}}|\times \hat{H}(\boldsymbol{k})|\boldsymbol{\partial_{k}} u_{n\boldsymbol{k}}\rangle$.

\subsection{Different gauge-removal schemes}\label{LC SR Gauge filtration}
The self-rotation and local circulation results become explicitly gauge-dependent once the previously omitted terms are incorporated.
We can interpret the gauge-independent self-rotation in Eq.~\eqref{SR in literatures} and local circulation in Eq.~\eqref{LC wrong 2} as outcomes obtained from the orbital moment operator by selectively eliminating gauge-dependent terms.
From this perspective, the discrepancy between the local circulation and the self-rotation arises from the difference in their gauge-removal schemes.
Here, we outline the gauge-removal processes used to obtain the self-rotation and local circulation results, and underscore their distinction from the gauge filtration devised in the main text.
We designate these conventional processes as the self-rotation and local circulation gauge-removal schemes, respectively.

First, we introduce the self-rotation and local circulation gauge-removal schemes.
The self-rotation result in Eq.~\eqref{SR in literatures} can be derived from Eq.~\eqref{SR correct} by either applying the gauge filtration proposed in the main text or substituting the position operator $\hat{\boldsymbol{r}}$ with a gauge-independent position operator $\hat{\boldsymbol{\mathcal{R}}}$.
We refer to this substitution of $\hat{\boldsymbol{r}}$ with $\hat{\boldsymbol{\mathcal{R}}}$ as the self-rotation gauge-removal scheme:
\begin{equation}
    \lim_{N\rightarrow \infty}
    \frac{1}{N}\int\frac{d\boldsymbol{k}}{V_{\text{B.Z.}}}\langle \Psi^{N}_{n\boldsymbol{k}}|\hat{\boldsymbol{\mathcal{R}}}\times\hat{\boldsymbol{v}}|\Psi^{N}_{n\boldsymbol{k}}\rangle
    =\int\frac{d\boldsymbol{k}}{V_{\text{B.Z.}}}
    \frac{i}{\hbar}\langle \boldsymbol{\partial_{k}}u_{n\boldsymbol{k}}|\times (\hat{H}(\boldsymbol{k})-E_{n}(\boldsymbol{k}))|\boldsymbol{\partial_{k}} u_{n\boldsymbol{k}}\rangle.
\end{equation}
Eliminating the surface term is necessary to obtain the local circulation result in Eq.~\eqref{LC wrong 2} from Eq.~\eqref{LC correct}.
We define this subtraction of the gauge-dependent surface term as the local circulation gauge-removal scheme.

The self-rotation and local circulation gauge-removal schemes appear to be viable alternatives for eliminating gauge dependence.
However, these schemes cannot be generalized when the operator $\hat{O}$ is combined with the orbital moment operator, such as $\{\hat{O},\hat{\boldsymbol{m}}\}_{+}$, because gauge-dependent terms arise that cannot be removed by either scheme alone.
In the self-rotation gauge-removal scheme, the orbital moment operator $\hat{\boldsymbol{m}}$ is substituted by $\hat{\boldsymbol{m}}^{\text{SR}}=(\hat{\boldsymbol{\mathcal{R}}}\times\hat{\boldsymbol{v}}-\hat{\boldsymbol{v}}\times\hat{\boldsymbol{\mathcal{R}}})/4$.
The evaluation of $\{\hat{O},\hat{\boldsymbol{m}}^{\text{SR}}\}_{+}$ inevitably contains gauge-dependent terms originating from the relation $i\boldsymbol{\partial_k}[O]_{nm}=[i\boldsymbol{\partial_{k}}O(\boldsymbol{k})]_{nm}-\big[\big\{[\boldsymbol{A}(\boldsymbol{k})],[O(\boldsymbol{k})]\big\}_{-}\big]_{nm}$:
\begin{equation}\label{SR GR scheme}
\begin{split}
    &\lim_{N\rightarrow \infty}\frac{1}{N}\sum_{\boldsymbol{R}}
    \langle \mathfrak{W}^{N}_{n\boldsymbol{R}}|\{\hat{O},\hat{\boldsymbol{m}}^{\text{SR}}\}_{+}|\mathfrak{W}^{N}_{n\boldsymbol{R}}\rangle
    =\frac{1}{4}\int \frac{d\boldsymbol{k}}{V_{\text{B.Z.}}}\bigg(
    \big([\{O(\boldsymbol{k}),\boldsymbol{A}(\boldsymbol{k})\times\boldsymbol{v}(\boldsymbol{k})-\boldsymbol{v}(\boldsymbol{k})\times\boldsymbol{A}(\boldsymbol{k})\}_{+}]_{nn}\big)^{\text{G.F.}}\\
    &\quad-\sum_{m}\Big(i\boldsymbol{\partial_k}[O(\boldsymbol{k})]_{nm}\times[\boldsymbol{v}(\boldsymbol{k})]_{mn}-[\boldsymbol{v}(\boldsymbol{k})]_{nm}\times i\boldsymbol{\partial_k}[O(\boldsymbol{k})]_{mn}\Big)
    \bigg).
\end{split}
\end{equation}
Explicitly expanding the gauge-dependent terms for a fixed $m$ in the summation yields:
\begin{equation}
\begin{split}
    &i\boldsymbol{\partial_k}[O(\boldsymbol{k})]_{nm}\times[\boldsymbol{v}(\boldsymbol{k})]_{mn}-[\boldsymbol{v}(\boldsymbol{k})]_{nm}\times i\boldsymbol{\partial_k}[O(\boldsymbol{k})]_{mn}\\
    &=[i\boldsymbol{\partial_{k}}O(\boldsymbol{k})]_{nm}\times[\boldsymbol{v}(\boldsymbol{k})]_{mn}-[\boldsymbol{v}(\boldsymbol{k})]_{nm}\times [i\boldsymbol{\partial_{k}}O(\boldsymbol{k})]_{mn}\\
    &-\Big[\big\{[\boldsymbol{A}(\boldsymbol{k})],[O(\boldsymbol{k})]\big\}_{-}\Big]_{nm}\times[\boldsymbol{v}(\boldsymbol{k})]_{mn}+[\boldsymbol{v}(\boldsymbol{k})]_{nm}\times\Big[\big\{[\boldsymbol{A}(\boldsymbol{k})],[O(\boldsymbol{k})]\big\}_{-}\Big]_{nm}\\
    &=[i\boldsymbol{\partial_{k}}O(\boldsymbol{k})]_{nm}\times[\boldsymbol{v}(\boldsymbol{k})]_{mn}-[\boldsymbol{v}(\boldsymbol{k})]_{nm}\times [i\boldsymbol{\partial_{k}}O(\boldsymbol{k})]_{mn}\\
    &+\sum_{l\neq n,m}\Big(
    \big([O(\boldsymbol{k})]_{nl}[\boldsymbol{A}(\boldsymbol{k})]_{lm}-[\boldsymbol{A}(\boldsymbol{k})]_{nl}[O(\boldsymbol{k})]_{lm}\big)\times[\boldsymbol{v}(\boldsymbol{k})]_{mn}\\
    &-
    [\boldsymbol{v}(\boldsymbol{k})]_{nm}\times\big([O(\boldsymbol{k})]_{ml}[\boldsymbol{A}(\boldsymbol{k})]_{ln}-[\boldsymbol{A}(\boldsymbol{k})]_{ml}[O(\boldsymbol{k})]_{ln}\big)
    \Big)\\
    &+\big([O(\boldsymbol{k})]_{nn}-[O(\boldsymbol{k})]_{mm}\big)\big([\boldsymbol{A}(\boldsymbol{k})]_{nm}\times[\boldsymbol{v}(\boldsymbol{k})]_{mn}+[\boldsymbol{v}(\boldsymbol{k})]_{nm}\times[\boldsymbol{A}(\boldsymbol{k})]_{mn}\big)\\
    &+\big([\boldsymbol{A}(\boldsymbol{k})]_{nn}-[\boldsymbol{A}(\boldsymbol{k})]_{mm}\big)\times\big([\boldsymbol{v}(\boldsymbol{k})]_{nm}[O(\boldsymbol{k})]_{mn}-[O(\boldsymbol{k})]_{nm}[\boldsymbol{v}(\boldsymbol{k})]_{mn}\big).
\end{split}
\end{equation}
This expression introduces an explicit gauge dependence due to the term $\big([\boldsymbol{A}(\boldsymbol{k})]_{nn}-[\boldsymbol{A}(\boldsymbol{k})]_{mm}\big)$ (where $m\neq n$) in the last line.
Crucially, this gauge dependence persists even after summing over $m$ in Eq.~\eqref{SR GR scheme}.

Similarly, we demonstrate that gauge-dependent terms remain after applying the local circulation gauge-removal scheme to the evaluation of $\{\hat{O},\hat{\boldsymbol{m}}\}_{+}$:
\begin{equation}\label{Om anticomm explicit}
\begin{split}
    &\lim_{N\rightarrow \infty}\frac{1}{N}\sum_{\boldsymbol{R}}\langle \mathfrak{W}^{N}_{n\boldsymbol{R}}|\{\hat{O},\hat{\boldsymbol{m}}\}_{+}|\mathfrak{W}^{N}_{n\boldsymbol{R}}\rangle
    =\lim_{N\rightarrow \infty}\frac{1}{N}\sum_{\boldsymbol{R}}\langle \mathfrak{W}^{N}_{n\boldsymbol{R}}|\{\hat{O},\hat{\boldsymbol{m}}^{\text{SR}}\}_{+}|\mathfrak{W}^{N}_{n\boldsymbol{R}}\rangle\\
    &+\frac{1}{4}\sum_{m}\int \frac{d\boldsymbol{k}}{V_{\text{B.Z.}}}
    \big([\boldsymbol{A}(\boldsymbol{k})]_{nn}+[\boldsymbol{A}(\boldsymbol{k})]_{mm}\big)\times\big([\boldsymbol{v}(\boldsymbol{k})]_{nm}[O(\boldsymbol{k})]_{mn}+[O(\boldsymbol{k})]_{nm}[\boldsymbol{v}(\boldsymbol{k})]_{mn}\big).
\end{split}
\end{equation}
To streamline our analysis, we focus on the gauge-dependent components within Eq.~\eqref{Om anticomm explicit}:
\begin{equation}\label{LC GR gauge dep terms}
\begin{split}
    &\frac{1}{2}\sum_{m}\int \frac{d\boldsymbol{k}}{V_{\text{B.Z.}}}
    \Big([\boldsymbol{A}(\boldsymbol{k})]_{nn}\times[\boldsymbol{v}(\boldsymbol{k})]_{nm}[O(\boldsymbol{k})]_{mn}
    +[O(\boldsymbol{k})]_{nm}[\boldsymbol{A}(\boldsymbol{k})]_{mm}\times [\boldsymbol{v}(\boldsymbol{k})]_{mn}\Big).
\end{split}
\end{equation}
Among these gauge-dependent terms, the local circulation gauge-removal scheme is exclusively applicable to the $m=n$ case, because a surface term to remove arises when integrating by parts to shift the derivative in the velocity expression $[\boldsymbol{v}(\boldsymbol{k})]_{nn}=\hbar^{-1}\boldsymbol{\partial_{k}}E_{n}(\boldsymbol{k})$.
Consequently, after applying the local circulation gauge-removal scheme, two distinct categories of gauge-dependent terms persist.
The first comprises the gauge-dependent terms for $m\neq n$ in Eq.~\eqref{LC GR gauge dep terms}.
The second category originates from the $m=n$ case in Eq.~\eqref{LC GR gauge dep terms}; i.e., for a given $n$:
\begin{equation}
\begin{split}
    &\int \frac{d\boldsymbol{k}}{V_{\text{B.Z.}}}[\boldsymbol{A}(\boldsymbol{k})]_{nn}[O(\boldsymbol{k})]_{nn}\times[\boldsymbol{v}(\boldsymbol{k})]_{nn}
    =\frac{1}{i\hbar}\int \frac{d\boldsymbol{k}}{V_{\text{B.Z.}}}[\boldsymbol{A}(\boldsymbol{k})]_{nn}[O(\boldsymbol{k})]_{nn}\times i\boldsymbol{\partial_{k}}E_{n}(\boldsymbol{k})\\
    &=\frac{1}{i\hbar}\int \frac{d\boldsymbol{k}}{V_{\text{B.Z.}}}E_{n}(\boldsymbol{k})i\boldsymbol{\partial_{k}} \times \big([\boldsymbol{A}(\boldsymbol{k})]_{nn}[O(\boldsymbol{k})]_{nn}\big)
    +\frac{1}{i\hbar}\int \frac{d\boldsymbol{k}}{V_{\text{B.Z.}}}i\boldsymbol{\partial_{k}} \times \big([\boldsymbol{A}(\boldsymbol{k})]_{nn}[O(\boldsymbol{k})]_{nn}E_{n}(\boldsymbol{k})\big).
\end{split}
\end{equation}
Within the first term in the last line, the contribution $\frac{1}{i\hbar}\int \frac{d\boldsymbol{k}}{V_{\text{B.Z.}}}E_{n}(\boldsymbol{k})\big(i\boldsymbol{\partial_{k}} [O(\boldsymbol{k})]_{nn}\big)\times [\boldsymbol{A}(\boldsymbol{k})]_{nn}$ remains explicitly dependent on the gauge choice, although the surface term (last term in the last line) is eliminated when the local circulation gauge-removal scheme is applied.

To address this limitation, we developed the gauge filtration method in the main text to systematically eliminate gauge dependence.
This framework inherently accounts for the Berry connection emerging in $i\boldsymbol{\partial_k}[O]_{nm}=[i\boldsymbol{\partial_{k}}O(\boldsymbol{k})]_{nm}-\big[\big\{[\boldsymbol{A}(\boldsymbol{k})],[O(\boldsymbol{k})]\big\}_{-}\big]_{nm}$.
Therefore, a gauge-removal rule tailored to a specific operator is not sufficiently general: even if the evaluation of the operator $\hat{\boldsymbol{m}}$ becomes gauge-independent via such a rule, the same rule cannot eliminate gauge dependence when $\hat{\boldsymbol{m}}$ is multiplied by other operators $\hat{O}$.

\section{Explicit calculations of the operators in the Wannier representation with the $\Delta_N$-regularization}\label{Wannier rep regularization}
This section details the explicit operator calculations referenced in Appendix~\ref{Regularization of Wannier functions}.
First, we demonstrate that $\lim_{N\rightarrow\infty}\langle \mathfrak{W}^{N}_{n'\boldsymbol{R}'}|\hat{O}|\mathfrak{W}^{N}_{n\boldsymbol{R}}\rangle=\langle W_{n'\boldsymbol{R}'}|\hat{O}|W_{n\boldsymbol{R}}\rangle$:
\begin{equation}\label{mathfrac W eval O}
\begin{split}
    &\lim_{N\rightarrow\infty}\langle \mathfrak{W}^{N}_{n'\boldsymbol{R}'}|\hat{O}|\mathfrak{W}^{N}_{n\boldsymbol{R}}\rangle
    =\iint \frac{d\boldsymbol{k}'d\boldsymbol{k}}{V_{\text{B.Z.}}^2}
    e^{i(\boldsymbol{k}'\cdot\boldsymbol{R}'-\boldsymbol{k}\cdot\boldsymbol{R})}
    \bigg(
    \lim_{N\rightarrow\infty}\langle\Psi^{N}_{n\boldsymbol{k}'}|\hat{O}|\Psi^{N}_{n\boldsymbol{k}}\rangle
    \bigg)\\
    &=\iint \frac{d\boldsymbol{k}'d\boldsymbol{k}}{V_{\text{B.Z.}}^2}
    e^{i(\boldsymbol{k}'\cdot\boldsymbol{R}'-\boldsymbol{k}\cdot\boldsymbol{R})}
    \bigg(
    \lim_{N\rightarrow\infty}V_{\text{B.Z.}}\int d\boldsymbol{k}''\Delta_{N}(\boldsymbol{k}'-\boldsymbol{k}'')[O(\boldsymbol{k}'')]_{n'n}\Delta_{N}(\boldsymbol{k}-\boldsymbol{k}'')
    \bigg)\\
    &=\iint \frac{d\boldsymbol{k}'d\boldsymbol{k}}{V_{\text{B.Z.}}^2}
    e^{i(\boldsymbol{k}'\cdot\boldsymbol{R}'-\boldsymbol{k}\cdot\boldsymbol{R})}
    \bigg(
    \lim_{N\rightarrow\infty}N\sum_{\boldsymbol{k}''}
    \delta_{\boldsymbol{k}'\boldsymbol{k}''}[O(\boldsymbol{k}'')]_{n'n}\delta_{\boldsymbol{k}\boldsymbol{k}''}
    \bigg)\\
    &=\iint \frac{d\boldsymbol{k}'d\boldsymbol{k}}{V_{\text{B.Z.}}^2}
    e^{i(\boldsymbol{k}'\cdot\boldsymbol{R}'-\boldsymbol{k}\cdot\boldsymbol{R})}
    \bigg(
    \lim_{N\rightarrow\infty}N\delta_{\boldsymbol{k}'\boldsymbol{k}}[O(\boldsymbol{k}')]_{n'n}
    \bigg)
    =\iint \frac{d\boldsymbol{k}'d\boldsymbol{k}}{V_{\text{B.Z.}}^2}
    e^{i(\boldsymbol{k}'\cdot\boldsymbol{R}'-\boldsymbol{k}\cdot\boldsymbol{R})}
    V_{\text{B.Z.}}\delta(\boldsymbol{k}'-\boldsymbol{k})[O(\boldsymbol{k}')]_{n'n}\\
    &=\int \frac{d\boldsymbol{k}}{V_{\text{B.Z.}}}
    e^{i\boldsymbol{k}\cdot(\boldsymbol{R}'-\boldsymbol{R})}[O(\boldsymbol{k})]_{n'n}
    =\langle W_{n'\boldsymbol{R}'}|\hat{O}|W_{n\boldsymbol{R}}\rangle.
\end{split}
\end{equation}

Second, we demonstrate that $\lim_{N\rightarrow\infty}N^{-1}\sum_{\tilde{\boldsymbol{R}}}\langle \mathfrak{W}^{N}_{n'(\boldsymbol{R}'+\tilde{\boldsymbol{R}})}|\hat{O}|\mathfrak{W}^{N}_{n(\boldsymbol{R}+\tilde{\boldsymbol{R}})}\rangle$ reduces to $N^{-1}\sum_{\tilde{\boldsymbol{R}}}\langle W_{n'(\boldsymbol{R}'+\tilde{\boldsymbol{R}})}|\hat{O}|W_{n(\boldsymbol{R}+\tilde{\boldsymbol{R}})}\rangle$, which, by virtue of cell periodicity, equals $\langle W_{n'\boldsymbol{R}'}|\hat{O}|W_{n\boldsymbol{R}}\rangle$:
\begin{equation}
\begin{split}
    &\lim_{N\rightarrow\infty}\frac{1}{N}\sum_{\tilde{\boldsymbol{R}}}\langle \mathfrak{W}^{N}_{n'(\boldsymbol{R}'+\tilde{\boldsymbol{R}})}|\hat{O}|\mathfrak{W}^{N}_{n(\boldsymbol{R}+\tilde{\boldsymbol{R}})}\rangle
    =\lim_{N\rightarrow\infty}\frac{1}{N}\sum_{\tilde{\boldsymbol{R}}}\iint \frac{d\boldsymbol{k}'d\boldsymbol{k}}{V_{\text{B.Z.}}^2}
    e^{i(\boldsymbol{k}'\cdot\boldsymbol{R}'-\boldsymbol{k}\cdot\boldsymbol{R})}e^{i(\boldsymbol{k}'-\boldsymbol{k})\cdot\tilde{\boldsymbol{R}}}
    \langle\Psi^{N}_{n\boldsymbol{k}'}|\hat{O}|\Psi^{N}_{n\boldsymbol{k}}\rangle\\
    &=\lim_{N\rightarrow\infty}\frac{1}{N}\iint \frac{d\boldsymbol{k}'d\boldsymbol{k}}{V_{\text{B.Z.}}}
    \delta(\boldsymbol{k}-\boldsymbol{k}')e^{i(\boldsymbol{k}'\cdot\boldsymbol{R}'-\boldsymbol{k}\cdot\boldsymbol{R})}
    \langle\Psi^{N}_{n'\boldsymbol{k}'}|\hat{O}|\Psi^{N}_{n\boldsymbol{k}}\rangle
    =\int \frac{d\boldsymbol{k}}{V_{\text{B.Z.}}}e^{i\boldsymbol{k}\cdot(\boldsymbol{R}'-\boldsymbol{R})}
    \bigg(
    \lim_{N\rightarrow\infty}\frac{1}{N}\langle\Psi^{N}_{n'\boldsymbol{k}}|\hat{O}|\Psi^{N}_{n\boldsymbol{k}}\rangle
    \bigg)\\
    &=\int \frac{d\boldsymbol{k}}{V_{\text{B.Z.}}}e^{i\boldsymbol{k}\cdot(\boldsymbol{R}'-\boldsymbol{R})}
    \bigg(
    \lim_{N\rightarrow\infty}\frac{V_{\text{B.Z.}}}{N}
    \int d\boldsymbol{k}''\Delta_{N}(\boldsymbol{k}-\boldsymbol{k}'')[O(\boldsymbol{k}'')]_{n'n}\Delta_{N}(\boldsymbol{k}-\boldsymbol{k}'')
    \bigg)\\
    &=\int \frac{d\boldsymbol{k}}{V_{\text{B.Z.}}}e^{i\boldsymbol{k}\cdot(\boldsymbol{R}'-\boldsymbol{R})}
    \bigg(
    \lim_{N\rightarrow\infty}\sum_{\boldsymbol{k}''}\delta_{\boldsymbol{k}\boldsymbol{k}''}[O(\boldsymbol{k}'')]_{n'n}\delta_{\boldsymbol{k}''\boldsymbol{k}}
    \bigg)
    =\int \frac{d\boldsymbol{k}}{V_{\text{B.Z.}}}
    e^{i\boldsymbol{k}\cdot(\boldsymbol{R}'-\boldsymbol{R})}[O(\boldsymbol{k})]_{n'n}\\
    &=\frac{1}{N}\sum_{\tilde{\boldsymbol{R}}}\langle W_{n'(\boldsymbol{R}'+\tilde{\boldsymbol{R}})}|\hat{O}|W_{n(\boldsymbol{R}+\tilde{\boldsymbol{R}})}\rangle
    =\langle W_{n'\boldsymbol{R}'}|\hat{O}|W_{n\boldsymbol{R}}\rangle.
\end{split}
\end{equation}

Third, we establish that $\lim_{N\rightarrow\infty}\langle \mathfrak{W}^{N}_{n'\boldsymbol{R}'}|\hat{\boldsymbol{r}}|\mathfrak{W}^{N}_{n\boldsymbol{R}}\rangle =\langle W_{n'\boldsymbol{R}'}|\hat{\boldsymbol{r}}|W_{n\boldsymbol{R}}\rangle=[\boldsymbol{A}(\boldsymbol{R}'-\boldsymbol{R})]_{n'n}+\boldsymbol{R}\delta_{n'n}\delta_{\boldsymbol{R}'\boldsymbol{R}}$:
\begin{equation}
\begin{split}
    &\lim_{N\rightarrow\infty}\langle \mathfrak{W}^{N}_{n'\boldsymbol{R}'}|\hat{\boldsymbol{r}}|\mathfrak{W}^{N}_{n\boldsymbol{R}}\rangle
    =\iint \frac{d\boldsymbol{k}'d\boldsymbol{k}}{V_{\text{B.Z.}}^2}
    e^{i(\boldsymbol{k}'\cdot\boldsymbol{R}'-\boldsymbol{k}\cdot\boldsymbol{R})}
    \bigg(
    \lim_{N\rightarrow\infty}\langle\Psi^{N}_{n\boldsymbol{k}'}|\hat{\boldsymbol{r}}|\Psi^{N}_{n\boldsymbol{k}}\rangle
    \bigg)\\
    &=\iint \frac{d\boldsymbol{k}'d\boldsymbol{k}}{V_{\text{B.Z.}}^2}
    e^{i(\boldsymbol{k}'\cdot\boldsymbol{R}'-\boldsymbol{k}\cdot\boldsymbol{R})}
    \bigg(
    \lim_{N\rightarrow\infty}V_{\text{B.Z.}}\int d\boldsymbol{k}''
    \Delta_{N}(\boldsymbol{k}'-\boldsymbol{k}'')[\boldsymbol{A}(\boldsymbol{k}'')]_{n'n}\Delta_{N}(\boldsymbol{k}-\boldsymbol{k}'')
    \bigg)
    +\iint \frac{d\boldsymbol{k}'d\boldsymbol{k}}{2V_{\text{B.Z.}}}
    e^{i(\boldsymbol{k}'\cdot\boldsymbol{R}'-\boldsymbol{k}\cdot\boldsymbol{R})}\\
    &\quad\times\lim_{N\rightarrow\infty}
    \delta_{n'n}\int d\boldsymbol{k}''
    \Big(
    \Delta_{N}(\boldsymbol{k}'-\boldsymbol{k}'')i\boldsymbol{\partial_{k''}}\Delta_{N}(\boldsymbol{k}-\boldsymbol{k}'')
    -\big(i\boldsymbol{\partial_{k''}}\Delta_{N}(\boldsymbol{k}'-\boldsymbol{k}'')\big)\Delta_{N}(\boldsymbol{k}-\boldsymbol{k}'')
    \Big)\\
    &=\int \frac{d\boldsymbol{k}}{V_{\text{B.Z.}}}e^{i\boldsymbol{k}\cdot(\boldsymbol{R}'-\boldsymbol{R})}
    \Big([\boldsymbol{A}(\boldsymbol{k})]_{n'n}+\boldsymbol{R}\delta_{n'n}\Big)
    =[\boldsymbol{A}(\boldsymbol{R}'-\boldsymbol{R})]_{n'n}+\boldsymbol{R}\delta_{n'n}\delta_{\boldsymbol{R}'\boldsymbol{R}},
\end{split}
\end{equation}
where the first term is calculated in the same manner as in Eq.~\eqref{mathfrac W eval O}, and the last term is derived using the relation
\begin{equation}
\begin{split}
    &\lim_{N\rightarrow\infty}\delta_{n'n}\iint \frac{d\boldsymbol{k}'d\boldsymbol{k}}{2V_{\text{B.Z.}}}
    e^{i(\boldsymbol{k}'\cdot\boldsymbol{R}'-\boldsymbol{k}\cdot\boldsymbol{R})}
    \int d\boldsymbol{k}''
    \Big(
    \Delta_{N}(\boldsymbol{k}'-\boldsymbol{k}'')i\boldsymbol{\partial_{k''}}\Delta_{N}(\boldsymbol{k}-\boldsymbol{k}'')
    -\big(i\boldsymbol{\partial_{k''}}\Delta_{N}(\boldsymbol{k}'-\boldsymbol{k}'')\big)\Delta_{N}(\boldsymbol{k}-\boldsymbol{k}'')
    \Big)\\
    &=\lim_{N\rightarrow\infty}\delta_{n'n}\iint \frac{d\boldsymbol{k}'d\boldsymbol{k}}{2V_{\text{B.Z.}}}
    e^{i(\boldsymbol{k}'\cdot\boldsymbol{R}'-\boldsymbol{k}\cdot\boldsymbol{R})}
    \int d\boldsymbol{k}''
    \Big(
    \big(i\boldsymbol{\partial_{k'}}\Delta_{N}(\boldsymbol{k}'-\boldsymbol{k}'')\big)\Delta_{N}(\boldsymbol{k}-\boldsymbol{k}'')
    -\Delta_{N}(\boldsymbol{k}'-\boldsymbol{k}'')i\boldsymbol{\partial_{k}}\Delta_{N}(\boldsymbol{k}-\boldsymbol{k}'')
    \Big)\\
    &=\delta_{n'n}\iint \frac{d\boldsymbol{k}'d\boldsymbol{k}}{2V_{\text{B.Z.}}}
    e^{i(\boldsymbol{k}'\cdot\boldsymbol{R}'-\boldsymbol{k}\cdot\boldsymbol{R})}
    \big(i\boldsymbol{\partial_{k'}}\delta(\boldsymbol{k}-\boldsymbol{k}')-i\boldsymbol{\partial_{k}}\delta(\boldsymbol{k}-\boldsymbol{k}')\big)
    =\delta_{n'n}\int \frac{d\boldsymbol{k}}{V_{\text{B.Z.}}}
    e^{i\boldsymbol{k}\cdot(\boldsymbol{R}'-\boldsymbol{R})}\frac{\boldsymbol{R}+\boldsymbol{R}'}{2}\\
    &=\frac{\boldsymbol{R}+\boldsymbol{R}'}{2}\delta_{n'n}\delta_{\boldsymbol{R}'\boldsymbol{R}}=\boldsymbol{R}\delta_{n'n}\delta_{\boldsymbol{R}'\boldsymbol{R}}.
\end{split}
\end{equation}

Finally, we verify that $\lim_{N\rightarrow\infty}N^{-1}\sum_{\tilde{\boldsymbol{R}}}\langle \mathfrak{W}^{N}_{n'(\boldsymbol{R}'+\tilde{\boldsymbol{R}})}|\hat{\boldsymbol{r}}|\mathfrak{W}^{N}_{n(\boldsymbol{R}+\tilde{\boldsymbol{R}})}\rangle =[\boldsymbol{A}(\boldsymbol{R}'-\boldsymbol{R})]_{n'n}$, which differs from $\lim_{N\rightarrow\infty}\langle \mathfrak{W}^{N}_{n'\boldsymbol{R}'}|\hat{\boldsymbol{r}}|\mathfrak{W}^{N}_{n\boldsymbol{R}}\rangle$ due to the absence of the term $\boldsymbol{R}\delta_{n'n}\delta_{\boldsymbol{R}'\boldsymbol{R}}$:
\begin{equation}\label{sum r in mathfrak W}
\begin{split}
    &\lim_{N\rightarrow\infty}\frac{1}{N}\sum_{\tilde{\boldsymbol{R}}}\langle \mathfrak{W}^{N}_{n'(\boldsymbol{R}'+\tilde{\boldsymbol{R}})}|\hat{\boldsymbol{r}}|\mathfrak{W}^{N}_{n(\boldsymbol{R}+\tilde{\boldsymbol{R}})}\rangle
    =\lim_{N\rightarrow\infty}\frac{1}{N}\sum_{\tilde{\boldsymbol{R}}}\iint \frac{d\boldsymbol{k}'d\boldsymbol{k}}{V_{\text{B.Z.}}^2}
    e^{i(\boldsymbol{k}'\cdot\boldsymbol{R}'-\boldsymbol{k}\cdot\boldsymbol{R})}e^{i(\boldsymbol{k}'-\boldsymbol{k})\cdot\tilde{\boldsymbol{R}}}
    \langle\Psi^{N}_{n'\boldsymbol{k}'}|\hat{\boldsymbol{r}}|\Psi^{N}_{n\boldsymbol{k}}\rangle\\
    &=\iint \frac{d\boldsymbol{k}'d\boldsymbol{k}}{V_{\text{B.Z.}}} \delta(\boldsymbol{k}-\boldsymbol{k}')
    e^{i(\boldsymbol{k}'\cdot\boldsymbol{R}'-\boldsymbol{k}\cdot\boldsymbol{R})}
    \bigg(
    \lim_{N\rightarrow\infty}\frac{1}{N}
    \langle\Psi^{N}_{n'\boldsymbol{k}'}|\hat{\boldsymbol{r}}|\Psi^{N}_{n\boldsymbol{k}}\rangle
    \bigg)
    =\int \frac{d\boldsymbol{k}}{V_{\text{B.Z.}}}e^{i\boldsymbol{k}\cdot(\boldsymbol{R}'-\boldsymbol{R})}
    \bigg(
    \lim_{N\rightarrow\infty}\frac{1}{N}
    \langle\Psi^{N}_{n'\boldsymbol{k}}|\hat{\boldsymbol{r}}|\Psi^{N}_{n\boldsymbol{k}}\rangle
    \bigg)\\
    &=\int \frac{d\boldsymbol{k}}{V_{\text{B.Z.}}}e^{i\boldsymbol{k}\cdot(\boldsymbol{R}'-\boldsymbol{R})}\bigg(\lim_{N\rightarrow\infty}\frac{V_{\text{B.Z.}}}{N}
    \int d\boldsymbol{k}''
    \Delta_{N}(\boldsymbol{k}-\boldsymbol{k}'')[\boldsymbol{A}(\boldsymbol{k}'')]_{n'n}\Delta_{N}(\boldsymbol{k}-\boldsymbol{k}'')
    \bigg)\\
    &+\int \frac{d\boldsymbol{k}}{V_{\text{B.Z.}}}e^{i\boldsymbol{k}\cdot(\boldsymbol{R}'-\boldsymbol{R})}\bigg(\lim_{N\rightarrow\infty}\frac{V_{\text{B.Z.}}}{N}
    \int d\boldsymbol{k}''\Delta_{N}(\boldsymbol{k}-\boldsymbol{k}'')\delta_{n'n}i\boldsymbol{\partial_{k''}}\Delta_{N}(\boldsymbol{k}-\boldsymbol{k}'')
    \bigg)\\
    &=\int d\boldsymbol{k}
    e^{i\boldsymbol{k}\cdot(\boldsymbol{R}'-\boldsymbol{R})}
    \bigg(
    \lim_{N\rightarrow\infty}\sum_{\boldsymbol{k}''}\delta_{\boldsymbol{k}\boldsymbol{k}''}[\boldsymbol{A}(\boldsymbol{k}'')]_{n'n}\delta_{\boldsymbol{k}''\boldsymbol{k}}
    \bigg)\\
    &=\int \frac{d\boldsymbol{k}}{V_{\text{B.Z.}}}
    e^{i\boldsymbol{k}\cdot(\boldsymbol{R}'-\boldsymbol{R})}[\boldsymbol{A}(\boldsymbol{k})]_{n'n}
    =[\boldsymbol{A}(\boldsymbol{R}'-\boldsymbol{R})]_{n'n}
    =\lim_{N\rightarrow\infty}\frac{1}{N}\sum_{\tilde{\boldsymbol{R}}}\langle \mathfrak{W}^{N}_{n'(\boldsymbol{R}'+\tilde{\boldsymbol{R}}+\tilde{\boldsymbol{R}}')}|\hat{\boldsymbol{r}}|\mathfrak{W}^{N}_{n(\boldsymbol{R}+\tilde{\boldsymbol{R}}+\tilde{\boldsymbol{R}}')}\rangle.
\end{split}
\end{equation}
To obtain this result, we utilized the identity:
\begin{equation}
\begin{split}
    &\int \frac{d\boldsymbol{k}}{V_{\text{B.Z.}}}e^{i\boldsymbol{k}\cdot(\boldsymbol{R}'-\boldsymbol{R})}
    \bigg(\lim_{N\rightarrow\infty}\frac{V_{\text{B.Z.}}}{N}
    \int d\boldsymbol{k}''\Big(\Delta_{N}(\boldsymbol{k}-\boldsymbol{k}'')\delta_{n'n}i\boldsymbol{\partial_{k''}}\Delta_{N}(\boldsymbol{k}-\boldsymbol{k}'')
    \bigg)\\
    &=\delta_{n'n}\int \frac{d\boldsymbol{k}}{V_{\text{B.Z.}}}e^{i\boldsymbol{k}\cdot(\boldsymbol{R}'-\boldsymbol{R})}
    \bigg(\lim_{N\rightarrow\infty}\frac{V_{\text{B.Z.}}}{2N}
    \int d\boldsymbol{k}'' i\boldsymbol{\partial_{k''}}\big(\Delta_{N}(\boldsymbol{k}-\boldsymbol{k}'')\big)^2
    \bigg)\\
    &=-\delta_{n'n}\int \frac{d\boldsymbol{k}}{V_{\text{B.Z.}}}e^{i\boldsymbol{k}\cdot(\boldsymbol{R}'-\boldsymbol{R})}
    \bigg(\lim_{N\rightarrow\infty}\frac{V_{\text{B.Z.}}}{2N}
    \int d\boldsymbol{k}'' i\boldsymbol{\partial_{k}}\big(\Delta_{N}(\boldsymbol{k}-\boldsymbol{k}'')\big)^2
    \bigg)\\
    &=-\delta_{n'n}\int \frac{d\boldsymbol{k}}{V_{\text{B.Z.}}}e^{i\boldsymbol{k}\cdot(\boldsymbol{R}'-\boldsymbol{R})}i\boldsymbol{\partial_{k}}
    \bigg(\lim_{N\rightarrow\infty}\frac{V_{\text{B.Z.}}}{2N}
    \int d\boldsymbol{k}'' \big(\Delta_{N}(\boldsymbol{k}-\boldsymbol{k}'')\big)^2
    \bigg)\\
    &=-(\boldsymbol{R}'-\boldsymbol{R})\delta_{n'n}\int \frac{d\boldsymbol{k}}{V_{\text{B.Z.}}}e^{i\boldsymbol{k}\cdot(\boldsymbol{R}'-\boldsymbol{R})}\frac{1}{2}
    =0,
\end{split}
\end{equation}
which directly implies $\frac{1}{2}\int d\boldsymbol{k}'' i\boldsymbol{\partial_{k''}}\bigg(\int \frac{d\boldsymbol{k}}{V_{\text{B.Z.}}}e^{i\boldsymbol{k}\cdot(\boldsymbol{R}'-\boldsymbol{R})}\big(\Delta_{N}(\boldsymbol{k}-\boldsymbol{k}'')\big)^{2}\bigg)=0$.

\end{widetext}
\end{appendix}
\end{document}